\documentclass[twocolumn, superscriptaddress, amsmath, amssymb, aps]{revtex4-2}

\usepackage{graphicx}
\usepackage{dcolumn}
\usepackage{bm}
\usepackage{hyperref}
\usepackage{siunitx}
\usepackage[normalem]{ulem}

\begin{document}

\preprint{APS/123-QED}

\title{
Likelihood Reconstruction for Radio Detectors of Neutrinos and Cosmic Rays
}

\author{Martin Ravn}
\affiliation{Dept. of Physics and Astronomy, Uppsala University, Box 516, S-75120 Uppsala, Sweden}

\author{Christian Glaser}
\affiliation{Dept. of Physics and Astronomy, Uppsala University, Box 516, S-75120 Uppsala, Sweden}
\affiliation{Dept. of Physics, TU Dortmund University, D-44221 Dortmund, Germany}

\author{Thorsten Glüsenkamp}
\affiliation{Dept. of Physics and Astronomy, Uppsala University, Box 516, S-75120 Uppsala, Sweden}
\affiliation{Oskar Klein Centre and Dept. of Physics, Stockholm University, SE-10691 Stockholm, Sweden}

\author{Ayca Özcelikkale}
\affiliation{Dept. of Electrical Engineering, Uppsala University, Box 65, 75103 Uppsala, Sweden}

\author{Alan Coleman}
\affiliation{Dept. of Physics and Astronomy, Uppsala University, Box 516, S-75120 Uppsala, Sweden}

\date{\today}

\begin{abstract}
    Ultra-high-energy neutrinos and cosmic rays are excellent probes of astroparticle physics phenomena. For astroparticle physics analyses, robust and accurate reconstruction of signal parameters such as arrival direction and energy is essential. Radio detection is an established detector concept explored by many observatories; however, current reconstruction methods ignore bin-to-bin noise correlations, which limits reconstruction resolution and, so far, has prevented calculations of event-by-event uncertainties. In this work, we present a likelihood description of neutrino or cosmic-ray signals in radio detectors with correlated noise, as present in all neutrino and cosmic-ray radio detectors. We demonstrate, with simulation studies of both neutrinos and cosmic-ray radio signals, that signal parameters such as energy and direction, including event-by-event uncertainties with correct coverage, can be obtained. This method reduces reconstruction uncertainties and biases compared to previous approaches. Additionally, the Likelihood can be used for event selection and enables differentiable end-to-end detector optimization. The reconstruction code is available through the open-source software NuRadioReco.
\end{abstract}

\maketitle

\section{Introduction}

Radio detection of ultra-high-energy (UHE) cosmic rays and neutrinos is a novel technique with the prospect of substantially advancing astroparticle physics in the coming years. For cosmic rays, the radio technique has been successfully used for many years at different air-shower observatories such as the Pierre Auger Observatory and LOFAR~\cite{Huege:2016veh, Schroder:2016hrv}. Its strengths include accurate energy estimation with small systematic uncertainties, as the emission can be calculated from first principles, insensitivity of the detector system to aging and environmental/atmospheric changes~\cite{PierreAuger:2015hbf,PierreAuger:2016vya, Glaser:2016qso, Gottowik:2017wio, Mulrey:2020oqe, PierreAuger:2023hun}, as well as precise measurement of the position of the shower maximum if the shower footprint is sampled with sufficient fidelity~\cite{Buitink:2014eqa,Buitink:2016nkf,PierreAuger:2023lkx,PierreAuger:2023rgk}. With the Auger-Prime upgrade, all surface detector stations of the Pierre Auger Observatory have been upgraded with radio antennas with the aim of measuring the chemical composition of cosmic rays at the highest energies~\cite{PierreAuger:2016qzd, PierreAuger:2023gql}. It is planned to enhance the IceTop air shower array of the IceCube detector at the South Pole with radio antennas~\cite{Haungs:2019ylq,IceCube-Gen2-TDR} and to equip the Square Kilometer Array (SKA) with cosmic-ray capabilities for precision air-shower physics~\cite{Buitink:2023rso}.

For neutrinos, the radio detection technique enables cost-effective instrumentation of large enough volumes to measure the faint flux of UHE neutrinos~\cite{Barwick:2022vqt} through ground-based air-shower arrays to detect Earth-skimming neutrinos~\cite{GRAND:2018iaj, beacon2020, TAROGE:2022soh}, balloon missions~\cite{ANITA:2008mzi, PUEO:2020bnn}, and through arrays of in-ice radio detector stations~\cite{RICE2003-Performance, ARA2011, ARIANNA:2019scz, RNO-G:2020rmc, IceCube-Gen2-TDR}. The latter, which will be the focus of this paper, was explored in the pilot arrays RICE~\cite{RICE2003-Performance, RICE2003-limits}, ARA~\cite{ARA2011, ARA2019-PA, ARA2020-limit, ARA:2022rwq}, and ARIANNA~\cite{ARIANNA:2019scz, Anker:2019rzo, Anker:2019zcx, ARIANNA:2020zrg, ARIANNA:2021pzm, Arianna:2021lnr, Arianna:2021vcx, Arianna:2023nvi} at the South Pole and Moore's Bay. With RNO-G~\cite{RNO-G:2020rmc}, a substantially larger detector, with the potential of measuring the first UHE neutrino, is currently being constructed in Greenland, and an order-of-magnitude more sensitive in-ice radio array is planned for IceCube-Gen2~\cite{IceCube-Gen2-TDR, IceCube-Gen2:2021rkf}.

All of these detectors have the same key challenge: Extracting the signal parameters from time series with low signal-to-noise ratios (SNRs). This key challenge will be addressed in this paper. We present, for the first time, a full likelihood reconstruction which promises to improve reconstruction performance. Furthermore, due to our correct probabilistic treatment, the likelihood not only estimates the true signal parameters but also provides the probability of any signal hypothesis, i.e., event-by-event uncertainties or the posterior probability density functions with arbitrary correlations between signal parameters. In the following paragraphs, we provide a brief summary of the current status of existing reconstruction methods before we present our work.

\subsection{Reconstructing radio signals from air showers}
Air shower arrays such as the Pierre Auger Observatory, LOFAR, or GRAND typically measure the radio signal on the ground with several dual-polarized antennas separated by $\mathcal{O}$(\SI{100}{m}) (an exception is LOFAR, which has a much denser sampling). Some antennas will have SNRs significantly above the noise floor, but due to the rapidly falling lateral signal distribution~\cite{Nelles:2014gma,Glaser:2018byo,Coleman:2023nra}, antennas further away from the shower axis will have small SNR. However, these antennas are crucial to determine the electromagnetic energy content and mass-sensitive parameters such as $X_\mathrm{max}$, and L/R (see e.g.~\cite{PierreAuger:2015hbf,Buitink:2014eqa,Corstanje:2023uyg}). So far, these analyses do not use the full information of the recorded time series but calculate a scalar variable, e.g., the energy fluence for two different polarization components of the signal. 

To improve on this, the \emph{forward folding} technique~\cite{Glaser:2017ctn, Glaser:2018ifj, Glaser:2019rxw} was developed, which uses a signal model whose predicted electric field is forward folded through the antenna and detector response and compared to the measured time series by subtracting the predicted from the measured signal in each time bin and adding the squares, i.e., calculating the chi-square ($\chi^2$). By minimizing this objective function, the best-fitting model parameters can be determined. It was found that the \emph{forward folding} method improved the reconstruction performance significantly at low SNRs~\cite{Glaser:2019rxw}. This method was successfully used to reconstruct air showers with the ARIANNA detector~\cite{Arianna:2021lnr}, and a simulation study predicted sensitivity to the cosmic-ray arrival direction and energy with a single detector station~\cite{Welling:2019scz}.

We note that the signal model is ideally an \emph{end-to-end} signal model that directly predicts the time series at each antenna for the parameters of interest such as cosmic-ray energy, direction, and $X_\mathrm{max}$. However, such a model is often difficult to formulate, has theoretical uncertainties, or is computationally very expensive. Alternatively, a more low-level reconstruction can be performed, where the electric field is reconstructed from each dual-polarized antenna, i.e., the signal arriving at the antennas. It turns out that the electric field can be parameterized analytically with high accuracy with just a few parameters, allowing a largely model-independent extraction of the signal parameters at each antenna location~\cite{Glaser:2019rxw}. 

\subsection{Reconstructing radio signals in dense media}
In-ice radio detectors for neutrinos are built as compact detector stations, each consisting of several (single polarization) antennas of different types with different orientations and positions with small separations of $\mathcal{O}$(\SI{10}{m}). By drilling holes in the ice, antennas can be lowered up to $\mathcal{O}$(\SI{100}{m}) beneath the surface, allowing for three-dimensional detector layouts. The stations are separated by $\mathcal{O}$(\SI{1}{km}) to observe mostly independent ice volumes and thus to maximize the total effective volume of the detector. Hence, each detector station must be able to \emph{independently} determine the relevant neutrino parameters. As a consequence, most observed neutrinos will have SNRs close to the trigger threshold. Even more, there are many ongoing efforts to further reduce trigger thresholds as much as possible, as an increased neutrino detection rate substantially improves the scientific outcome~\cite{ARA2019-PA, Glaser:2020pot, Arianna:2021vcx, Glaser:2023RO}. Hence, extracting the signal parameters at low SNRs is of the utmost importance. 

The station design of in-ice detectors includes several single polarization antennas with $\mathcal{O}$(\SI{1}{m} -- \SI{10}{m}) spacing and thus each antenna observes a different electric field. Thus, unlike for the dual-polarized antennas that are used in air shower arrays, a reconstruction of the electric field per antenna is not uniquely defined. Hence, to use the \emph{forward folding} technique, a model is needed that simultaneously predicts the signal in all antennas. Fortunately, this is readily available via parameterizations or semi-analytic calculations of the \emph{Askaryan emission} (see~\cite{Glaser:2019cws} for an overview of the available models) that predict the electric field for any observer position (relative to the in-ice particle shower), shower energy, and interaction type. Effects from propagating the signal through ice are added, after which the signal is folded with the antenna and detector response to obtain the time series that every antenna is expected to observe. Simulation studies show that the \emph{forward folding} technique can be used to estimate the neutrino direction~\cite{Gaswint:2021smu, ARIANNA:2021pzm,Plaisier:2023cxz, IceCube-Gen2:2023czw} and it achieves the best resolution of all traditional methods, i.e., not considering deep learning.

The forward folding method has been shown to reliably obtain best-fit parameters of radio neutrino and cosmic-ray signals; however, published implementations of the method have not developed a complete likelihood description and have therefore not been able to provide reliable event-by-event uncertainties. The main reason for this is that the noise observed in radio detectors has strong correlations between neighboring bins, but these correlations are ignored in current reconstruction methods. For instance, in the case of the forward folding method, when calculating the difference between data and model prediction bin-by-bin, the sum of the differences squared divided by the variance of the noise can not be interpreted as a $\chi^2$. The main missing ingredient in properly accounting for these correlations and obtaining event-by-event uncertainties is a probabilistic description of the noise observed in radio detectors, which we will develop in this paper. In general, a measured trace consists of a signal plus one realization of probabilistic noise. A probabilistic description of the noise will then allow us to calculate the likelihood of a signal, given the measured trace. Using the likelihood in a reconstruction is the most robust and powerful way of extracting the signal parameters from a measurement if the signal model can be fully specified. 

We note that in practice, it can be challenging to fully specify the signal model and its dependence on the parameters of interest, especially for non-deterministic cases such as high-energy electron neutrino interactions where the LPM effect introduces significant shower-to-shower fluctuations~\cite{Landau:1953um,Landau:1965ksp,Migdal:1956tc}. In such cases, better performance might be achievable with deep learning. Either only the signal model is parameterized via neural networks and used in a likelihood fit (as successfully done by IceCube to discover neutrino emission from the Galactic plane~\cite{IceCube:2023ame}), or a deep neural network is directly trained to predict the parameters of interest~\cite{Glaser:2022lky}. By incorporating conditional Normalizing Flows into neural networks, the network can predict the probability of any model parameter including correlations, just as in the likelihood method~\cite{Glusenkamp:2020gtr} (which was recently demonstrated for IceCube~\cite{IceCube:2023avo} and in-ice radio detectors~\cite{IceCube-Gen2:2023wcw}). In this paper, we do not consider deep learning. 

In this paper, we present a probabilistic description of the noise observed in radio neutrino or cosmic ray detectors. The model can be used to express a likelihood for a neutrino or cosmic-ray radio signal, given a set of measured traces in an arbitrary number of antennas, which can be used in a reconstruction of signal parameters. In Sec.~\ref{sec_noise_model}, we describe the probabilistic noise model and demonstrate its compatibility with simulated data. In Sec.~\ref{sec_applications}, we demonstrate the various applications of the noise model: neutrino signal reconstruction for an in-ice radio detector (Sec.~\ref{sec_neutrino_reconstruction}), fast uncertainty estimation for differentiable end-to-end detector optimization (Sec.~\ref{sec_fisher_information_matrix}), reconstruction of electric fields from cosmic ray air showers with a dual-polarized antenna (Sec.~\ref{sec_electric_field_reconstruction}), and identification of cosmic-ray signals in an in-ice radio detector (Sec.~\ref{sec_cosmic_ray_identification}).

\section{Probabilistic noise model}  \label{sec_noise_model}

In this section, we describe the noise present in radio detectors and propose how to describe it probabilistically. We discuss various theoretical aspects of the proposed model and verify it against simulated noise data. 

\subsection{Noise in radio detectors of cosmic rays and neutrinos} \label{sec_radio_noise}
The noise present in all radio detectors falls into two main categories: anthropogenic noise and noise natural to the environment or the detector. Anthropogenic noise consists primarily of spurious pulsed signals. They are, in general, considered to be an identifiable background and can be differentiated from the target physics signals in event selections. Additionally, many modern radio detectors are built in remote places that are sufficiently radio-quiet that noise pulses are rare and unlikely to overlap with the neutrino and cosmic-ray signals that only last a few nanoseconds. Therefore, we do not need to model pulsed noise for event reconstruction. We note that pulsed noise is still a challenge for self-triggering and identifying the few cosmic rays and neutrinos from all triggered events. Only the latter is addressed in this paper. Anthropogenic noise from radio broadcasting and communication can form continuous signals in detectors; however, these often span very narrow frequency ranges or single frequencies and can be filtered out in the frequency domain, e.g., by notch filters.

In contrast, natural noise sources form a diffuse stochastic background that is continuously present in the detector across all frequencies. The main contributions are thermal radio noise from the environment, radio noise from the Milky Way galaxy~\cite{international1982recommendations}, and noise from the detector electronics themselves. The environmental thermal noise and galactic noise are observed through the antennas and are hence folded with the antenna response, whereas the electronic noise is added in various steps in the signal chain and is heavily dependent on the hardware, e.g., amplifiers (see e.g.~\cite{Mulrey:2019vtz}). In total, these sources form a continuous background that can not be eliminated but is well understood and can be characterized.

Noise in radio detectors of cosmic rays and neutrinos has been studied extensively in the past. After basic data cleaning (e.g., high- and low-pass filtering, removal of frequency bins contaminated with continuous wave signals), the noise can be described as bandwidth-limited Gaussian noise. A procedure was developed to simulate this noise in the Fourier domain (see, e.g., the implementation in NuRadioReco~\cite{Glaser:2019rxw}), which can be transformed back into the time domain via an inverse Fourier transform. Studies of noise from several radio neutrino experiments show Rayleigh-distributed frequency amplitudes (equivalent to our assumption of bandwidth-limited Gaussian noise, see below) \cite{Hong:2014vbg, Meures:2014rfw, ANITA:2019xbi, ARA:2024def}. In addition, multiple analyses have shown excellent agreement between measurements and simulations, e.g.,~\cite{Arianna:2021vcx, Arianna:2023nvi}, where a deep neural network trained on simulated data performed as expected on experimental data from the ARIANNA detector, providing indirect evidence that the noise has been simulated correctly.
We describe the noise generation method and our notation in the following.

\subsection{Notation and formalism} \label{sec_formalism}

A radio detector measures the voltages in a number of antennas as a function of time, which we denote $V(t_n)$. The data is sampled at discrete times $t_n$, called bins or samples, and recorded in traces of $n_t$ consecutive measurements. We have $t_n =t_0 +n \times \delta t$,  for $n=0, \ldots, n_t-1$ where $\delta t$ is the time between the samples, and  $T= n_t  \times \delta t$ is the duration of the trace. In the derivation below, it is assumed that $n_t$ is even and $t_0$ is set as $t_0=0$ for convenience. The discrete Fourier transform of a trace $V(t_0), \ldots, V(t_{n_t-1})$ is given by
\begin{equation} \label{eq_rfft}
	\mathcal{V}(\omega_k)  =  \sqrt{2} \, \delta t \, \left( \sum_{n=0}^{n_t-1}  V(t_n) e^{- i \omega_k t_n} \right),
\end{equation}
where $\mathcal{V}(\omega_k)$  is the complex amplitude of the  Fourier mode corresponding to the frequency $\omega_k = k \frac{2\pi}{T}$, where $-(\frac{n_t}{2} -1) \leq k \leq \frac{n_t}{2}$. Since $V(t)$  is real-valued, there is Hermitian symmetry in the Fourier spectrum, i.e., $\mathcal{V}(\omega_k) = \mathcal{V}^*(-\omega_k)$ where $^*$ denotes complex conjugate. Hence, we only need to calculate $\mathcal{V}(\omega_k)$ from $k=0$ to $k=n_t/2$, i.e., the Nyquist frequency. The trace can be recovered from the spectrum using the following inverse discrete Fourier transform
\begin{equation} \label{eq_inv_rfft_general} 
\begin{aligned} 
	V(t_n) = & \frac{1}{\sqrt{2} \, \delta t} \cdot \frac{1}{n_t} \, \sum_{k= -(n_t/2-1)}^{n_t/2}  \mathcal{V}(\omega_k) e^{i\omega_k t_n} \\
     = &\frac{1}{\sqrt{2} \, \delta t} \cdot \frac{1}{n_t} \, \bigg(  \mathcal{V}(\omega_0) e^{i\omega_0 t_n} + \mathcal{V}(\omega_{n_t/2}) e^{i\omega_{n_t/2} t_n} \\ & + 2 \sum_{k= 1}^{n_t/2-1}  \mathfrak{Re} \{\mathcal{V}(\omega_k) e^{i\omega_k t_n} \} \bigg),  
\end{aligned}
\end{equation}
where we used Hermitian symmetry to represent $V(t_n)$ using only the non-negative frequencies.

The normalization factor $\sqrt{2} \, \delta t$ in Eq.~\eqref{eq_rfft} ensures that the units in the frequency domain is voltage per frequency, and that the energy is conserved under reasonable assumptions that approximately hold in practice in a typical scenario.  In particular, we consider the following conditions: 
(A1) there is no bias in the noise, i.e., the average value of  the noise trace (when there is no signal present) is zero so that $\frac{1}{n_t}\sum_{n=0}^{n_t-1}  V(t_n) =0$, hence  $\mathcal{V}(0) = 0$; 
(A2) $\mathcal{V}(\omega_{n_t/2}) =0$, i.e., the sampling rate is sufficiently high so that the non-zero part of the noise spectrum is captured by sampling, and hence the contribution from the Fourier mode at the Nyquist frequency is zero.
Then, the energy is conserved as follows:  
\begin{equation} \label{eq_power} 
	\sum_{n=0}^{n_t-1} |V(t_n)|^2 \, \delta t = \sum_{k=0}^{n_t/2} |\mathcal{V}(\omega_k)|^2 \, \delta f,
\end{equation}
where $\delta f = \frac{1}{n_t \delta t}$ is the length of the frequency bins. Under (A1) and (A2), Eq.~\eqref{eq_inv_rfft_general} can be conveniently re-written as
\begin{equation} \label{eq_inv_rfft}
\begin{aligned}
	V(t_n) = \frac{\sqrt{2}}{n_t \, \delta t} \sum_{k=1}^{n_t/2-1} \Big[ \mathfrak{Re}\{\mathcal{V}(\omega_k)\} \cos(\omega_k t_n)& \\- \mathfrak{Im}\{\mathcal{V}(\omega_k)\} \sin(\omega_k t_n) &\Big].
\end{aligned}
\end{equation}

The noise generation procedure is then as follows: First, a real-valued spectrum $A(\omega_k)$ is chosen for the noise generation process. This spectrum can either be a theoretical spectrum or be measured directly from data. Second, the noise realization is generated in the frequency domain by pulling the real and imaginary parts of $\mathcal{V}(\omega_k)$ from Gaussian distributions with zero mean and equal variance of $\frac{1}{2} (A(\omega_k))^2$, statistically independent over the frequency index $k$, and over real and imaginary parts for a given $k$. We note that for $k=0$ and $k=n_t/2$, purely real  $\mathcal{V}(\omega_k)$ is formed using a Gaussian distribution with variance $(A(\omega_k))^2$ to satisfy the condition $V(t_n)$ is real-valued. This procedure results in Rayleigh-distributed amplitudes and uniformly distributed phases for $\mathcal{V}(\omega_k)$ for $1 \leq k \leq \frac{n_t}{2}-1$. The relationship between $A(\omega_k)$ and the expectations of the Fourier mode amplitudes is given by $\text{E}[|\mathcal{V}(\omega_k)|^2] = A(\omega_k)^2$.  Finally, the noise is inverse Fourier transformed to the time domain according to Eq.~\eqref{eq_inv_rfft_general}, or equivalently according to Eq.~\eqref{eq_inv_rfft} if $A(\omega_0 ) =0$ and $A(\omega_{n_t/2} ) =0$. This formalism of generating noise is the established standard in the \emph{in-ice radio} community, the same as that used in NuRadioMC~\cite{Glaser:2019cws}, and it was verified numerous times to describe experimentally measured noise (see e.g.~\cite{Arianna:2021vcx,Arianna:2023nvi}).

\begin{figure*}[t]
	\centering
	\includegraphics[width=1\textwidth,trim={0cm 0.3cm 0cm 0cm},clip]{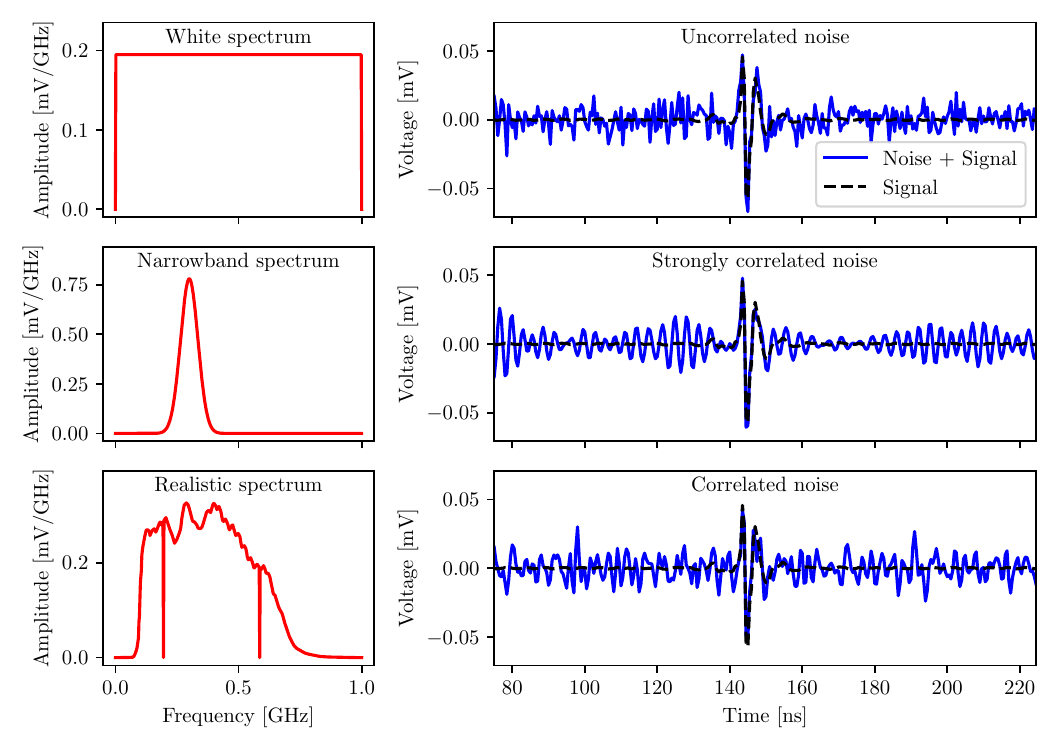}
	\caption{Different noise spectra and resulting time domain noise realizations with an injected simulated neutrino signal. The realistic spectrum is based on the averaged noise spectrum from a single RNO-G antenna presented in Ref.~\cite{Agarwal:20259H}, to which we added two notch filters.}
	\label{fig_noise}
\end{figure*}

If the spectrum is chosen to be a box function extending from 0 to the Nyquist frequency, the generated noise will be uncorrelated Gaussian noise, also called white noise. If, however, a band-limited spectrum with a cut-off at low and/or high frequencies is chosen, the generated noise will have correlations between neighboring samples. The latter is the case for the noise observed in radio detectors. We show a few examples of different noise scenarios in Fig.~\ref{fig_noise} with an injected simulated neutrino signal. The first row shows white noise, the second row shows unrealistically strongly band-limited noise, and the last row shows a realistic noise spectrum. The realistic spectrum is based on the averaged noise spectrum for a single RNO-G antenna presented in Figure~3 in Ref.~\cite{Agarwal:20259H}. Additionally, we added two notch filters to demonstrate the robustness of our method in cases where anthropogenic continuous-wave (CW) signals need to be filtered out. We note that the method presented in this section works for any noise spectrum with Rayleigh-distributed amplitudes, as long as the spectrum is known or can be measured.

In any problem where successive measurements are Gaussian and uncorrelated, the standard objective function to minimize in order to fit a curve, e.g., $\mu(t_n; \boldsymbol{\theta})$, parameterized by $\boldsymbol{\theta} = (\theta_0,  \ldots, \theta_{n_\theta})$, to a measured dataset, e.g., $V(t_n)$, is the chi-square,
\begin{equation} \label{eq_chi2}
    \chi^2 \equiv \sum_{n=0}^{n_t-1} \frac{[V(t_n) - \mu(t_n; \boldsymbol{\theta})]^2}{\sigma^2},
\end{equation}
where $\sigma$ is the standard deviation of the noise. However, if this objective function is used for bandwidth-limited noise, all information about correlations is ignored, and it will not be chi-square distributed for repeated measurements. It can therefore not be interpreted statistically like a chi-square, and it can not be used to estimate uncertainties on reconstructed parameters. This is the reason why previous work on reconstruction using the forward folding technique, i.e., a chi-square reconstruction, did not predict any event-by-event uncertainties~\cite{Glaser:2019rxw, Gaswint:2021smu, ARIANNA:2021pzm, Arianna:2021lnr, Plaisier:2023cxz, IceCube-Gen2:2023czw}. To obtain an objective function that can be interpreted statistically, the correct probabilistic description of band-limited noise is needed. We present such a model in this paper, which allows us to calculate the likelihood of a signal in a radio detector given a measured trace.

\subsection{Probabilistic noise model} \label{sec_probabilistic_noise_model}

We model the noise plus a deterministic signal in a radio detector as a multivariate normal distribution in $n_t$ dimensions. The mean of each bin is the signal, and the deviation is a random number from a normal distribution correlated with the neighboring bins. By representing the trace and signal as vectors, $\boldsymbol{x} = (V(t_0), \ldots, V(t_{n_t-1}))^\mathrm{T}$ and $\boldsymbol{\mu}(\boldsymbol{\theta}) = (\mu(t_0;\boldsymbol{\theta}), \ldots, \mu(t_{n_t-1};\boldsymbol{\theta}))^\mathrm{T}$, the probability density function for measuring a trace given a signal is then:
\begin{equation} \label{eq_multivariate_normal}
\begin{aligned}
	p\big(\boldsymbol{x};\boldsymbol{\mu}&(\boldsymbol{\theta}),\boldsymbol{\Sigma}\big) = \frac{1}{\sqrt{(2\pi)^{n_t} |{\boldsymbol{\Sigma}}| }} \\ & \times \exp \bigg(-\frac{1}{2}\big(\boldsymbol{x}-\boldsymbol{\mu}(\boldsymbol{\theta})\big)^\mathrm{T} \boldsymbol{\Sigma}^{-1} \big(\boldsymbol{x}-\boldsymbol{\mu}(\boldsymbol{\theta})\big) \bigg),
\end{aligned}
\end{equation}
where $\boldsymbol{\Sigma}$ is the covariance matrix of the noise and $|{\boldsymbol{\Sigma}}|$ is its determinant. Here, it is assumed that $\boldsymbol{\Sigma}$ is invertible, whereas the non-invertible case is discussed in Sec.~\ref{sec_pseudo_inverse}. The covariance matrix parameterizes the correlations between the noise in different time bins, and we discuss how to determine it in the following sections. The multivariate normal distribution hence describes correlated noise and can be used to calculate the probability that a trace belongs to the expected distribution, potentially including a neutrino or cosmic ray signal.

The probability density function in Eq.~\eqref{eq_multivariate_normal} can then be seen as the likelihood for a signal given a measured trace. For statistical analyses, minus two times the natural logarithm of the likelihood is of special significance,
\begin{widetext}
\begin{equation} \label{eq_llh}
	\begin{aligned}
		-2\ln \mathcal{L}(\boldsymbol{\mu}(\boldsymbol{\theta});\boldsymbol{x},\boldsymbol{\Sigma})
        =  n_t \ln(2 \pi) + n_t \ln(|{\boldsymbol{\Sigma}}|)  + \big(\boldsymbol{x}-\boldsymbol{\mu}(\boldsymbol{\theta})\big)^\mathrm{T} \boldsymbol{\Sigma}^{-1}  \big(\boldsymbol{x}-\boldsymbol{\mu}(\boldsymbol{\theta})\big).
	\end{aligned}
\end{equation}
\end{widetext}
This is the ideal objective function to minimize in a reconstruction, and we denote the maximum likelihood estimate of the signal parameters as $\hat{\boldsymbol{\theta}} = \text{arg min}_{\boldsymbol{\theta}} \big(-2\ln{\mathcal{L}(\boldsymbol{\mu}(\boldsymbol{\theta});\boldsymbol{x},\boldsymbol{\Sigma})}\big)$. For this purpose, the first two constant terms can be ignored, and the log-likelihood can easily be summed over the different channels of a radio detector if the signal is present in several antennas. Additionally, this objective function, and differences in the objective function between two hypotheses, 
\begin{equation}
	\begin{aligned}
		-2\Delta\ln{\mathcal{L}(\boldsymbol{\theta}_1, \boldsymbol{\theta}_2)} = -2[\ln \mathcal{L}(\boldsymbol{\mu}&(\boldsymbol{\theta}_2);\boldsymbol{x},\boldsymbol{\Sigma}) 
        \\ & - \ln \mathcal{L}(\boldsymbol{\mu}(\boldsymbol{\theta}_1);\boldsymbol{x},\boldsymbol{\Sigma}]),
	\end{aligned}
\end{equation}
can be interpreted statistically since they are chi-square distributed according to Wilks' theorem~\cite{Wilks:1938dza}. It can hence be used to derive event-by-event uncertainties on the reconstructed parameters and estimate a goodness-of-fit for the reconstructed signal. The main goal of this paper is to demonstrate that the likelihood can be used for reconstructing signals across the field of radio detection with correct event-by-event uncertainties and that it improves upon the reconstruction uncertainties compared to previous methods.

In the following, we will discuss how to calculate the covariance matrix and its inverse and show the model's compatibility with simulated noise. All aspects discussed here are implemented in a noise model class available in NuRadioReco, which includes methods for calculating the likelihood for any radio signal with Gaussian noise.

\subsection{The covariance matrix} \label{sec_covariance_matrix}

The covariance of measurements at two different times, $t_i$ and $t_j$, is defined as:
\begin{equation} \label{eq_covariance_ij}
	\mathrm{Cov}(t_i,t_j) = \mathrm{E}[(V(t_i)-\mu_i)(V(t_j)-\mu_j)],
\end{equation}
where $\mathrm{E}[\cdot]$ denotes the expectation, i.e., statistical average, operator and $\mu_{i} =\mathrm{E}[V(t_i)]$ and $\mu_{j}=\mathrm{E}[V(t_j)] $ are the averages of the measurements at $t_i$ and $t_j$. In practice, the expectations, $\widehat{\mu}_{i}$ and $\widehat{\mu}_{j}$, can be calculated as the average over $N$ independent realizations of traces containing only noise, either measured or simulated. Hence, the empirical covariance values can be estimated numerically from the traces as follows:
\begin{equation} \label{eq_covariance_empirical}
    \widehat{\mathrm{Cov}}(t_i,t_j) = \frac{1}{N}\sum_{m=1}^{N} (x_{m,i}-\widehat{\mu}_i)(x_{m,j}-\widehat{\mu}_j),
\end{equation}
where $x_{m,i}$ and $x_{m,j}$ are the $m^{\rm{th}}$ realizations or measurement of $V(t_{i})$ and $V(t_{j})$, and $\widehat{\mu}_{i}$ and $\widehat{\mu}_{j}$ are expected to be zero for traces containing purely noise.

When $i=j$ the covariance reduces to the variance, $\mathrm{Cov}(t_i,t_i) = \mathrm{Var}(t_i)$. The covariance matrix, $\boldsymbol{\Sigma}$, then has entries $\Sigma_{ij} = \mathrm{Cov}(t_i,t_i)$ and has the structure:
\begin{equation}
\begin{aligned}
    &\boldsymbol{\Sigma} = \\&
    \begin{pmatrix} 
		\mathrm{Var}(t_0) & \mathrm{Cov}(t_1,t_0) & \hdots & \mathrm{Cov}(t_{n_t-1},t_0) \\ 
		\mathrm{Cov}(t_0,t_1) & \mathrm{Var}(t_1) & \hdots & \mathrm{Cov}(t_{n_t-1},t_1) \\ 
		\vdots & \vdots & \ddots & \vdots \\
		\mathrm{Cov}(t_0,t_{n_t-1}) & \mathrm{Cov}(t_1,t_{n_t-1}) & \hdots & \mathrm{Var}(t_{n_t-1}) \\ 
    \end{pmatrix} .
\end{aligned}
\end{equation}
Note that $\mathrm{Cov}(t_i,t_j)= \mathrm{Cov}(t_j,t_i)$, $\forall t_i,t_j$, hence the covariance matrices are always symmetric.

\begin{figure*}[t]
	\includegraphics[width=1\textwidth,trim={0 0.4cm 0 0.3cm},clip]{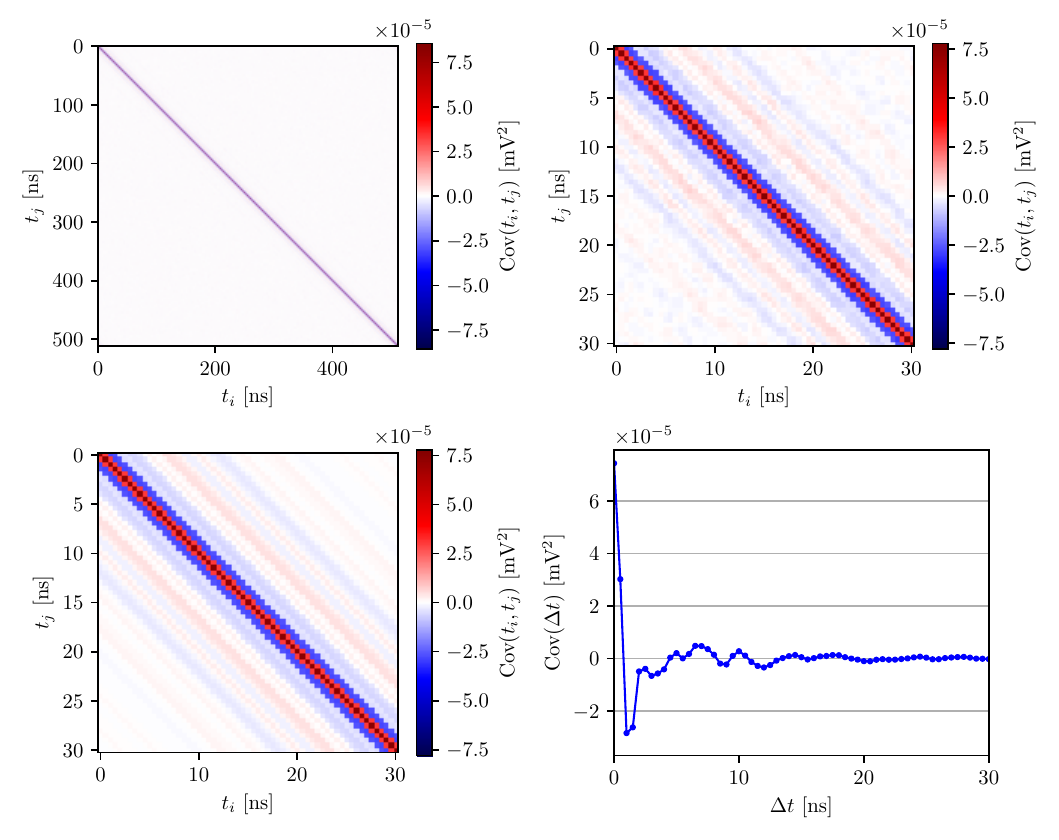}
	\caption{\textit{Top left}: The covariance matrix calculated empirically from 10,000 traces of generated noise using the realistic spectrum shown in Fig.~\ref{fig_noise}. \textit{Top right}: Zoom in on the first \SI{30}{ns} by \SI{30}{ns} of the covariance matrix. \textit{Bottom left}: Covariance matrix calculated empirically from 10,000 traces of generated noise averaged over the diagonals according to a symmetric circulant structure (zoom in on the first \SI{30}{ns} by \SI{30}{ns}). \textit{Bottom right}: The values of the diagonals of the averaged covariance matrix as a function of $\Delta t = t_j - t_i$.}
	\label{fig_covariance_matrix}
\end{figure*}

For the noise we are considering here, the variance is assumed to be constant within the time period under consideration. Hence, the diagonal of the covariance matrix is constant. For band-limited noise, measurements close to each other (small $\Delta t = t_j - t_i$) are expected to be strongly correlated, and the correlations should decrease for longer time intervals. The entries of the covariance matrix close to the diagonal should thus be large and decrease going away from the diagonal. 
In Fig.~\ref{fig_covariance_matrix} (Top left and Top right), we show an example of a typical covariance matrix. We calculated the covariance matrix according to Eq.~\eqref{eq_covariance_empirical} from 2,000 traces of simulated band-limited thermal noise using the realistic spectrum shown in Fig.~\ref{fig_noise}. The left panel shows the full covariance matrix, and the right panel shows a zoom-in of the relevant features close to the diagonal.

We assume that the correlation between the values of noise at two time instants solely depends on how close these two instants are, i.e, $\mathrm{Cov}(t_i,t_j)$ only depends on $|t_i-t_j|$. As a result of this translation invariance in time and the fact that the traces are obtained using equidistant samples, the covariance matrix has a symmetric Toeplitz structure. In particular, the values on the diagonals are constant, and the matrix is symmetric around the main diagonal. A symmetric Toeplitz matrix has the following structure:
\begin{equation}
	\bm{T}_\text{s} = \begin{pmatrix} 
		c_0 & c_{1} & \hdots & c_{n_t-2} & c_{n_t-1} \\ 
		c_{1} & c_0 & c_{1} & \hdots & c_{n_t-2} \\ 
		\vdots & c_1 & c_0 & \ddots & \vdots \\
		c_{n_t-2} & \vdots & \ddots & \ddots & c_{1} \\ 
		c_{n_t-1} & c_{n_t-2} & \hdots & c_1 & c_0 \\ 	
	\end{pmatrix}.
\end{equation}
When constructing an estimate of the covariance matrix from traces, we need to make assumptions about the structure of the covariance matrix. The symmetric Toeplitz matrix assumption is attractive due to both the intuitively appealing translational invariance and also due to the fact that the statistical uncertainty on the estimate of the covariance matrix can be reduced by averaging over the diagonals. 

We can make another simplification by observing that the formalism to generate noise (as described above) will result in periodic signals, i.e., the last time bins are correlated with the first time bins of the trace. This is because we represent a real-valued finite-duration sampled time series using a finite number of modes in the Fourier domain. Although this periodicity is unphysical and not present in data, it is not of any practical concern because of the following reasons. 
The bin-to-bin correlations of noise are local, i.e., they only span over a few time bins as we will show below. Hence, ignoring the first and last few time bins will remove the unphysical correlations. Furthermore, the expected neutrino and cosmic-ray signals are short and well contained within a trace. Therefore, it is common practice in a typical reconstruction pipeline to switch between representing the measurement in the time domain and the Fourier domain, depending on what is more appropriate for the analysis step~\cite{Glaser:2019rxw}. Any potential problem can be further mitigated by filtering the edges of the trace in the time domain to zero, e.g., using a Hanning window, and prepending/appending the traces with zeros. Hence, representing and modifying signals in the Fourier domain using a finite number of Fourier modes, which may implicitly make the trace cyclic, is not a practical problem, but simplifies many processing steps, and we will adopt it in the following.

Hence, the structure of the covariance matrix can be further restricted as a symmetrical circulant matrix, which is a special case of the Toeplitz matrix. It has the structure:
\begin{equation}
	\bm{C}_\text{s} = \begin{pmatrix} 
		c_0 & c_1 & \hdots & c_2 & c_1 \\ 
		c_1 & c_0 & c_1 & \hdots & c_2 \\ 
		\vdots & c_1 & c_0 & \ddots & \vdots \\
		c_2 & \vdots & \ddots & \ddots & c_1 \\ 
		c_1 & c_2 & \hdots & c_1 & c_0 \\ 	
	\end{pmatrix},
\end{equation}
The symmetric circulant matrix has half the number of parameters as the Toeplitz matrix, and the statistical uncertainty of each entry is thus decreased.

Assuming that the covariance matrix is circulant mainly affects its upper right and lower left corners, where the correlations between the first and last few bins are described. We experience that for long traces, the decreased statistical uncertainty outweighs any negative effects from assuming periodicity. In this paper, we thus use the symmetrical circulant matrix to describe the covariance matrix. A zoomed in part of the covariance matrix averaged over the diagonals according to the circulant structure is shown in Fig.~\ref{fig_covariance_matrix} (Bottom left). Due to the symmetry of the matrix, the covariance matrix can be represented in 1D as the value of the covariance as a function of the distance to the diagonal, i.e., the time difference between the two bins. We show this representation in Fig.~\ref{fig_covariance_matrix} (Bottom right). We observe what also corresponds to our intuition that the variance ($\Delta t = 0$) is the largest value and that the magnitude of the correlation quickly decays with increasing time separation. We also see that neighboring bins have a positive correlation, but for slightly larger time separation, the bins are negatively correlated.

\subsection{Inversion of the covariance matrix and the pseudoinverse} \label{sec_pseudo_inverse}
In order to evaluate the multivariate normal probability of a trace given a signal (Eq.~\eqref{eq_multivariate_normal}), the inverse of the covariance matrix is required. In cases where the covariance matrix is of full rank, the inverse can simply be calculated with standard inversion techniques. However, this is often not the case, and its inverse, hence, does not exist. Here, we describe the origin of the problem and how it can be solved. 

In essentially all realistic applications, some frequency amplitudes are zero (or numerically so close to zero that they can be treated as being exactly zero). This comes in part from the antenna response and the filters built into the signal chain of the detector, as well as additional digital processing where frequency filters are often applied to data to remove unwanted frequencies. This could, for instance, be high and low-pass filters to remove regions of low SNR, and notch filters to remove anthropogenic CW signals. Every complex-valued frequency bin filtered out removes two degrees of freedom in the time domain. Hence, every removed frequency subtracts two degrees of freedom from a trace in the time domain, and the degrees of freedom are $n_{\rm{dof}} = n_t - 2\, n_{f,\rm{removed}}$. Since the rank of the covariance matrix is equal to the degrees of freedom, the covariance matrix will not be of full rank if frequency filtering is applied, which describes a degenerate distribution. The covariance matrix is therefore not invertible, and the probability density function in Eq.~\eqref{eq_multivariate_normal} cannot be evaluated.

The correct way to describe a degenerate multivariate normal distribution is to replace the inverse covariance matrix with the Moore-Penrose pseudoinverse $\boldsymbol{\Sigma}^{-1} \rightarrow \boldsymbol{\Sigma}^{+}$, and the determinant with the pseudo-determinant $|{\boldsymbol{\Sigma}}| \rightarrow |{\boldsymbol{\Sigma}}|_+$~\cite{rao1973linear}. The Moore–Penrose pseudoinverse of a covariance matrix is computed by first performing the eigendecomposition $\boldsymbol{\Sigma} = \boldsymbol{Q} \boldsymbol{\Lambda} \boldsymbol{Q}^{-1}$, where $\boldsymbol{\Lambda} = \mathrm{diag}(\lambda_1, \, \dots, \, \lambda_{n_t})$ is a diagonal matrix comprised of the eigenvalues of $\boldsymbol{\Sigma}$. By inverting all non-zero entries of the diagonal matrix, $\boldsymbol{\Lambda}^+ = \mathrm{diag}(\lambda_1^+, \, \dots, \, \lambda_{n_t}^+)$, where $\lambda_k^+ = 1/\lambda_k \text{ if } \lambda_k > 0 \text{ and } 0 \text{ if } \lambda_k =0$, the Moore-Penrose pseudoinverse is obtained, $\boldsymbol{\Sigma}^+ = \boldsymbol{Q} \boldsymbol{\Lambda}^+ \boldsymbol{Q}^{-1}$. The pseudo-determinant is then the product of all non-zero eigenvalues, $|{\boldsymbol{\Sigma}}|_+ = \prod_{\lambda_k > 0} \lambda_k$. In numerical implementation, the Moore–Penrose pseudoinverse and its determinant are typically computed using a threshold $\epsilon$, where eigenvalues below this value are regarded as zero to ensure numerical stability. The degenerate multivariate normal distribution can then be expressed as
\begin{equation} \label{eq_degenerate_normal}
\begin{aligned}
	p\big(\boldsymbol{x};\boldsymbol{\mu}&(\boldsymbol{\theta}),\boldsymbol{\Sigma}\big) = \frac{1}{\sqrt{(2\pi)^{n_\text{dof}} |{\boldsymbol{\Sigma}}|_+ }} \\ & \times \exp \bigg(-\frac{1}{2}(\boldsymbol{x}-\boldsymbol{\mu}(\boldsymbol{\theta}))^\mathrm{T} \boldsymbol{\Sigma}^{+} (\boldsymbol{x}-\boldsymbol{\mu}(\boldsymbol{\theta})) \bigg),
\end{aligned}
\end{equation}
where $n_{\text{dof}} = n_t - 2n_{f,\text{removed}}$. The Moore-Penrose pseudoinverse and determinant exist and are unique for all matrices, and they coincide with the regular inverse and determinant if the covariance matrix is of full rank. Throughout this work, we use the notation of the regular and pseudoinverse interchangeably and apply the Moore-Penrose pseudoinverse where needed.

An additional benefit of using the pseudoinverse of the covariance is that it improves the numerical stability of evaluating the probability density function. If numerical errors are added to a point from the distribution, it may no longer lie exactly in the hyperplane where the distribution exists, and the likelihood can thus not be evaluated. However, when using the pseudoinverse, the width of the distribution is infinite in the filtered-out dimensions, and it is thus not sensitive to numerical errors.

\subsection{Analytical expression of covariance matrix and its inverse}\label{sec_analytical_method}
Under the noise generation process described after Eq.~\eqref{eq_inv_rfft} and assuming $A(\omega_0 ) =0$ and $A(\omega_{n_t/2} ) =0$,
the covariance of the two measurements $V(t_i)$ and $V(t_j)$ can be shown to have the following form:
\begin{equation}\label{eq_analytical_covariance}
	\begin{aligned}
    	\mathrm{Cov}(t_i,t_j) 
      = & \frac{2}{n_t^2  \, \delta t^2} \frac{1}{2}\sum_{k=1}^{n_t/2-1} A(\omega_k)^2 \cos(\omega_k \Delta t_{i,j}), 
     \end{aligned}
\end{equation}
where $\Delta t_{i,j} = t_j - t_i$. We see here that the covariance matrix depends only on $\Delta t$, i.e., the distance from the diagonal and not $t_i$ or $t_j$ alone, and is symmetrical around both $\Delta t = 0$ (the diagonal) and $\Delta t = T/2$. It is thus a symmetrical circulant matrix as discussed in Sec.~\ref{sec_covariance_matrix}. 

\begin{figure*}[t]
	\centering
	\includegraphics[width=1\textwidth,trim={0 0.4cm 0 0.3cm},clip]{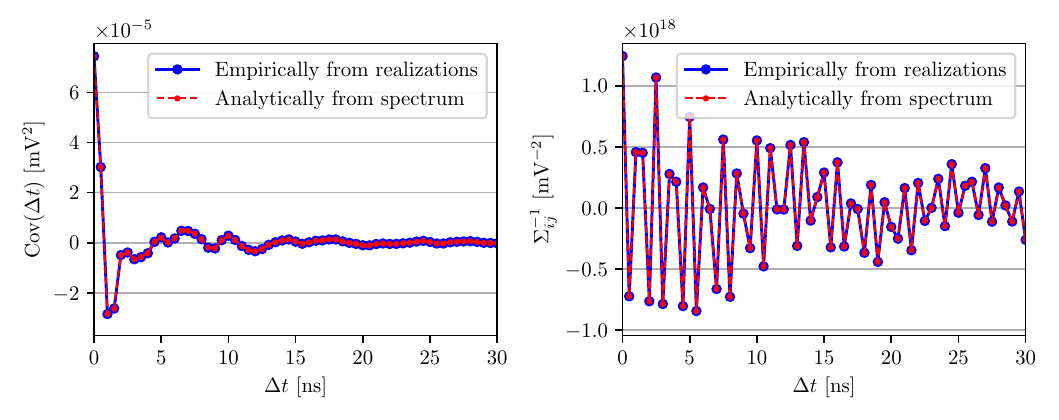}
	\caption{One row of the covariance matrix (left) and its inverse (right) for simulated noise calculated empirically from 10,000 traces of noise using Eq.~\eqref{eq_covariance_ij} and analytically from the spectrum using Eq.~\eqref{eq_analytical_covariance}. The difference between the two calculations is less than 4‰ of the element on the diagonal ($\Delta t = 0$), which we attribute to the statistical uncertainties of calculating the covariance matrix from a limited number of noise realizations.}
	\label{fig_cov_analytical}
\end{figure*}

To find the inverse of the covariance matrix, Eq.~\eqref{eq_analytical_covariance} can be expressed as
\begin{equation}\label{eq_analytical_covariance_2}
	\begin{aligned}
        \boldsymbol{\Sigma} = 
        & \boldsymbol{F}^\text{H} \text{diag}\bigg(\frac{1}{2 n_t \delta t^2} A'(\omega_{k'})^2 \bigg) \boldsymbol{F}, 
     \end{aligned}
\end{equation}
where $\boldsymbol{F}$ is the discrete Fourier transform matrix with elements $F_{nk'} = \frac{1}{\sqrt{n_t}} e^{i \omega_{k'} \Delta t_{0,n}}$, $\boldsymbol{F}^\text{H} = \boldsymbol{F}^{-1}$ is its Hermitian transpose and inverse, and we let $k'$ go from $0$ to $n_t-1$. The different scaling of the Fourier transform here compared to Eq.~\eqref{eq_rfft} ensures that the column vectors of $\boldsymbol{F}^\text{H}$ form a normalized basis. The matrix $\text{diag}\Big(\frac{1}{2 n_t \, \delta t^2} A'(\omega_k)^2\Big)$ is a $n_t \times n_t$ diagonal matrix and $A'(\omega_{k'})$ are the amplitudes defined for $0 \leq k' \leq n_t-1$ with $A'(\omega_{k'}) = A'(\omega_{n_t - k'}) = A(\omega_k)$. Since $\boldsymbol{F}^\text{H}$ is the matrix with normalized column vectors that diagonalizes $\boldsymbol{\Sigma}$, the columns of $\boldsymbol{F}^\text{H}$ are the eigenvectors and the entries of the diagonal matrix are the eigenvalues of $\boldsymbol{\Sigma}$. We observe that the covariance matrix's rank (number of non-zero eigenvalues) is equal to two times the number of non-zero frequency amplitudes, $\text{rank}(\boldsymbol{\Sigma}) = 2 n_{A(\omega_k)>0}$. Additionally, the inverse covariance matrix is
\begin{equation}\label{eq_analytical_inverse}
	\begin{aligned}
        \boldsymbol{\Sigma}^{-1} = 
        & \boldsymbol{F}^\text{H} \text{diag}\bigg(\frac{2 n_t \delta t^2}{A'(\omega_{k'})^2} \bigg) \boldsymbol{F},
     \end{aligned}
\end{equation}
or
\begin{equation}\label{eq_analytical_inverse_2}
	\begin{aligned}
    	\boldsymbol{\Sigma}^{-1}_{ij} 
      = & \frac{2}{n_t} \sum_{k=1}^{n_t/2-1} \frac{2 n_t \delta t^2}{A(\omega_k)^2} \cos(\omega_k \Delta t_{i,j}).
     \end{aligned}
\end{equation}
From this expression, we see that if any of the amplitudes are equal to zero, the inverse of the covariance matrix is not defined, which is consistent with the discussion in Sec.~\ref{sec_pseudo_inverse}. When this is the case, the sum is calculated only for $A(\omega_k)>0$, which gives the Moore-Penrose pseudoinverse of the covariance matrix. In numerical implementations, a non-zero but small threshold amplitude is set to avoid numerical uncertainties, and the sum is calculated for $A(\omega_k)>A_{\text{threshold}}$. Additionally, one can choose to ignore frequency components of signals by omitting them from the sum above.

Eq.~\eqref{eq_analytical_covariance} and \eqref{eq_analytical_inverse} provides an analytical way of calculating the covariance matrix of noise and its inverse directly from the spectrum. One row of the covariance matrix and one row of its inverse calculated analytically from the spectrum is shown in Fig.~\ref{fig_cov_analytical} alongside the empirical calculation from many noise realizations using Eq.~\eqref{eq_covariance_ij}. The two results are visually indistinct from each other and agree well. As the average noise spectrum is easily obtainable from data or can be calculated from measurements of the signal chain (S21 parameter), using the analytical expressions presented in this section is a robust and convenient way to calculate the covariance matrix and its inverse, and will be used in the following.

The development discussed here naturally leads to a frequency domain formulation of the multivariate normal distribution, which is presented in Appendix~\ref{sec_frequency_domain}. This expression is numerically faster to evaluate and more stable. It will hence be employed in the following. Furthermore, the relation between the likelihood and \emph{matched filtering} is discussed in Appendix~\ref{sec_matched_filter}. The matched filtering approach will be used as part of a reconstruction algorithm (see Sec.~\ref{sec_forward_folding_likelihood}) to improve stability and for efficient and optimal signal identification (see Sec.~\ref{sec_cosmic_ray_identification}).

\subsection{Verification of noise model}
\label{sec_verification}

To verify that the multivariate normal distribution describes noise in radio detectors well, we evaluate its consistency with noise generated by the established and well-tested procedure described in Sec.~\ref{sec_formalism}. This is done by calculating the minus two log probability for many realizations of noise:
\begin{equation} \label{eq_log_prob}
	\begin{aligned}
		-2 \ln p(\boldsymbol{x}&; \boldsymbol{\mu}(\boldsymbol{\theta}),\boldsymbol{\Sigma})  \\ & = \big(\boldsymbol{x}-\boldsymbol{\mu}(\boldsymbol{\theta})\big)^\mathrm{T} \boldsymbol{\Sigma}^{-1}  \big(\boldsymbol{x}-\boldsymbol{\mu}(\boldsymbol{\theta})\big) + norm.
	\end{aligned}
\end{equation}
If the normalization is ignored, Eq.~\eqref{eq_log_prob} is equivalent to a chi-square with correlations taken into account. It should hence be chi-square distributed with degrees of freedom equal to the number of degrees of freedom of the multivariate normal distribution, if the distribution describes the noise correctly. If all frequency amplitudes are non-zero and the covariance matrix is full rank, we have $n_\text{dof} = n_t$; otherwise, the degrees of freedom are equal to two times the number of non-zero frequency amplitudes or, equivalently, the rank of the (inverse) covariance matrix.

\begin{figure*}[!htb]
 	\centering
 	\includegraphics[width=1\textwidth,trim={0 0.4cm 0 0.3cm},clip]{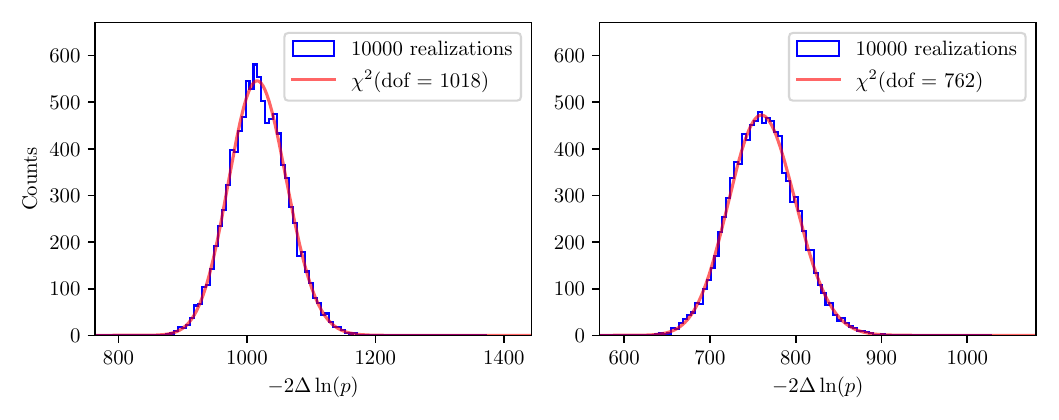}
 	\caption{$-2 \Delta \ln{p}$ distributions for 10,000 realizations of simulated noise using the realistic spectrum in Fig.~\ref{fig_noise}. The distributions have been calculated with the inverse covariance matrix (Eq.~\eqref{eq_analytical_inverse_2}) using the full spectrum except the first, the last, and the two filtered out frequency bins (left) and ignoring all frequency amplitudes below 1\% of the maximal amplitude of the spectrum (right). Alongside the distributions are chi-square distributions with the expected degrees of freedom for the two cases shown.}
 	\label{fig_llh_distributions}
 \end{figure*}
 
In Fig.~\ref{fig_llh_distributions}, the $-2 \Delta \ln{p}$ (normalization subtracted) distribution is shown for 10,000 realizations of noise using the realistic spectrum shown in Fig.~\ref{fig_noise}. The resulting distributions are shown for two different cases: First (left), using the full spectrum except for the first, the last, and the two filtered out frequency bins, which gives $n_\text{dof} = n_\text{samples}-6 = 1018$ degrees of freedom. Second (right), ignoring all frequency amplitudes below 1\% of the maximal amplitude which results in $n_\text{dof} = \text{rank}(\boldsymbol{\Sigma^+}) = 762$. The inverse covariance matrices were calculated analytically from the spectrum using Eq.~\eqref{eq_analytical_inverse_2}, and the calculations were repeated with the empirically determined covariance matrices, giving the same result within numerical uncertainties. The figure shows perfect agreement between the $-2 \Delta \ln{p}$ distributions of the noise realization and the chi-square distribution with the corresponding degrees of freedom. This experiment again illustrates that Eq.~\eqref{eq_analytical_covariance} provides the covariance values for noise generated according to Sec.~\ref{sec_formalism}, and the resulting time-domain variables are Gaussian distributed as expected.

\section{Applications} \label{sec_applications}

The probabilistic noise model and the corresponding likelihood have vast applicability and could substantially improve the analysis of radio-based detection of cosmic rays and neutrinos. This is because all detectors record bandwidth-limited signals, i.e., the noise is correlated, and all detectors need to extract the signal properties from low signal-to-noise ratio data, as discussed in the introduction. In the following, we quantify the impact of our new model on several typical reconstruction and event selection tasks.

\subsection{Reconstruction of neutrinos for in-ice radio detectors} \label{sec_neutrino_reconstruction}
With a probabilistic description of noise in radio detectors, it is possible to perform a maximum-likelihood reconstruction of neutrino signals in in-ice radio detectors. In this section, we demonstrate with a Monte-Carlo (MC) simulation study of a radio detector station foreseen for the IceCube-Gen2 radio array~\cite{IceCube-Gen2-TDR}, that using the likelihood description presented in this paper, signal parameters can be accurately reconstructed with correct uncertainties. Additionally, we study the improvements in reconstruction performance compared to previous work, which used an uncorrelated $\chi^2$ (Eq.~\eqref{eq_chi2}, e.g., used in ~\cite{Glaser:2019rxw, Gaswint:2021smu,Plaisier:2023cxz}).

A preliminary study has been presented in Ref.~\cite{Ravn:2024arx}. In this paper, we improve the realism of the study by using the detector layout proposed for IceCube-Gen2 radio and taking into account the bending of radio signals in ice due to a depth-dependent refractive index.

\subsubsection{Neutrino signal simulation} \label{sec_neutrino_signal_model}

All simulations of neutrino signals described here are performed with NuRadioMC~\cite{Glaser:2019cws}. The Askaryan radio emission~\cite{Askaryan:1961pfb,1965JETP...21..658A} resulting from the time-varying charge excess of a particle shower initiated by a neutrino can be calculated analytically in the frequency domain. We employ one such parametrization, which is referred to as \textit{Alvarez2009}~\cite{Alvarez-Muniz:2005mez, Alvarez-Muniz:2009jsq, Alvarez-Muniz:2010hbb}. The parametrization provides an efficient way to calculate a realistic neutrino signal and is well-suited for a simulation-based reconstruction study. In this study, we consider only the so-called \emph{hadronic shower}, i.e., in-ice cascades initiated by the breakup of the nucleus through Z or W exchange with the incoming neutrino. Due to the many initial hadronic interactions, these cascades have low shower-to-shower fluctuations, and the emitted signal can hence be considered deterministic, unlike electron-induced showers from electron-neutrino charged-current interactions, which have large stochasticities due to the LPM effect (see, e.g.,~\cite{Glaser:2019cws, Coleman:2024scd}).

\begin{figure}[t]
    \centering
    \includegraphics[width=0.8\linewidth]{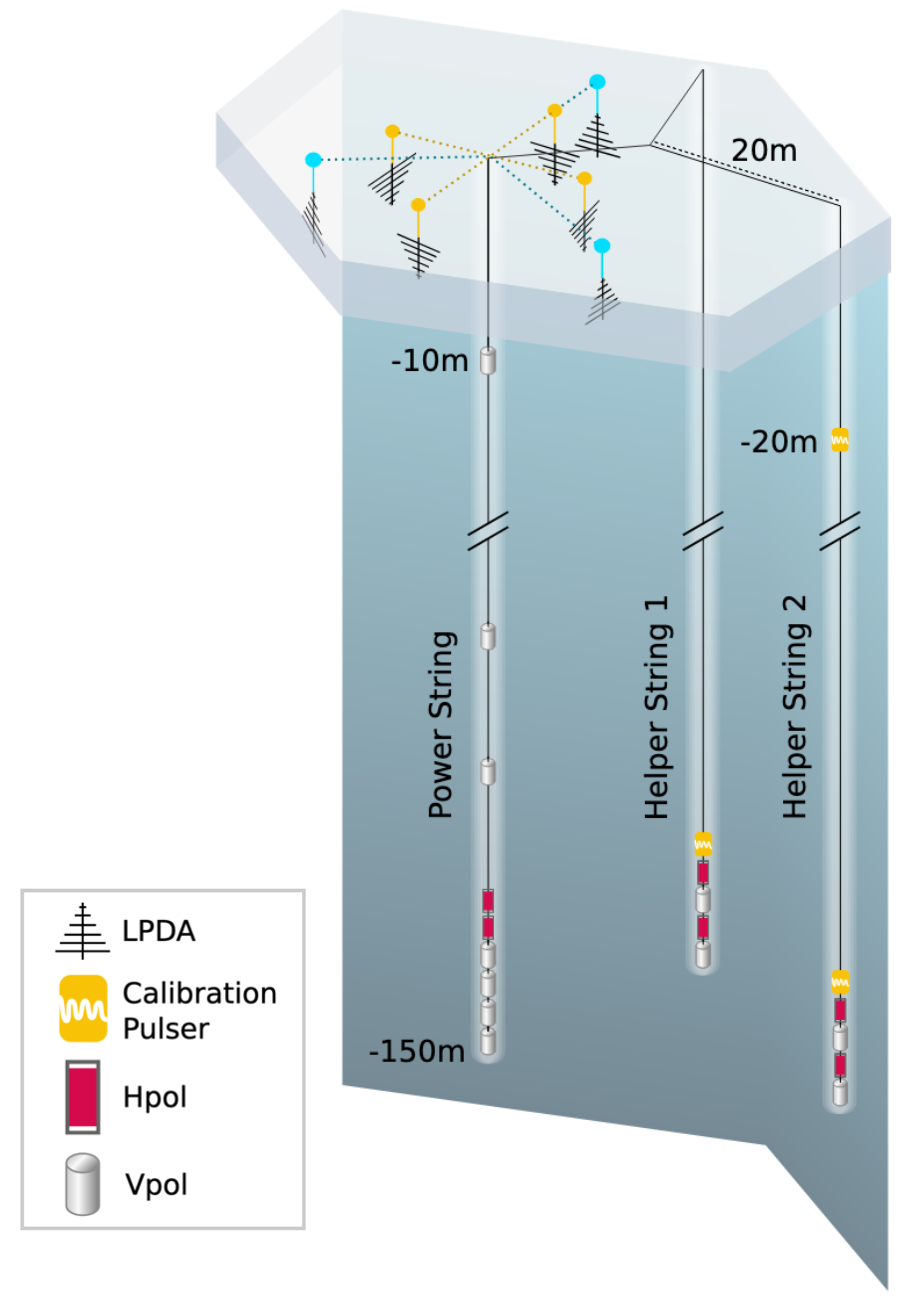}
    \caption{Sketch of the planned IceCube-Gen2 hybrid station radio detector layout. Figure adapted from Ref.~\cite{IceCube-Gen2-TDR}.}
    \label{fig_gen2}
\end{figure}

We approximate the index-of-refraction profile of the ice with an exponential function, which was used in previous studies and provides a good description of measured data~\cite{Barwick:2018rsp}. The likelihood reconstruction will work with arbitrary ice models, but we choose this simple model as it provides a good trade-off between speed and accuracy for this proof-of-concept study. 

The resulting radio neutrino signal depends on seven parameters: the energy of the particle shower ($E_\text{shower}$), the neutrino arrival direction (zenith angle $\theta_\nu$ and azimuth angle $\phi_\nu$), the neutrino vertex position (expressed in spherical coordinates through zenith angle $\theta_\text{vertex}$, azimuth angle $\phi_\text{vertex}$, and distance $r_\text{vertex}$), and an overall time offset ($t_0$). The neutrino vertex position in spherical coordinates is defined in accordance with NuRadioMC, i.e., relative to an observer located at $x=y=z=0$. These seven parameters are the ones we aim to estimate in a reconstruction, where neutrino energy and arrival direction are of special interest for astrophysical analyses.

The detector simulation is performed using NuRadioReco~\cite{Glaser:2019rxw}. It provides a useful data structure for events and a wide range of antenna and signal chain descriptions. With NuRadioReco, the detector response to an electrical field, as calculated by NuRadioMC, can be simulated, and a trace for a neutrino signal in a given antenna and position is obtained.

\begin{figure*}[t]
	\centering
    \includegraphics[width=1\textwidth,trim={0 0.4cm 0 0.2cm},clip]{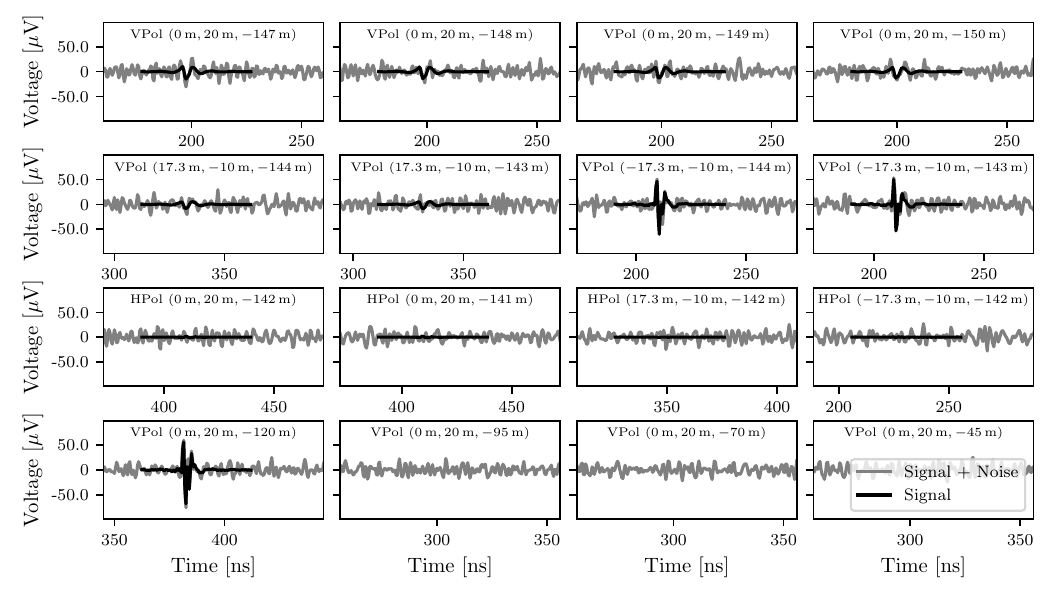}
	\caption{Simulated neutrino signal in 16 antennas with one realization of noise for Event 3 in Table~\ref{tab_events}. The panels show the types and $(x, \, y, \, z)$ coordinates of the antennas, and the signals fall outside the readout windows in the last three panels.}
	\label{fig_signal}
\end{figure*}

This reconstruction study is performed using the station layout foreseen for the IceCube-Gen2 radio array, which is shown in Fig.~\ref{fig_gen2} Specifically, we study events in the deep part of the detector, which consists of 16 antennas located on strings in 3 boreholes. At the bottom of the main string, called the \textit{power string}, at \SI{150}{m} depth, four vertically polarized (VPol) antennas are located with a vertical distance of \SI{1}{m}. These antennas act as a phased array and trigger the station~\cite{Allison:2018ynt}. Directly above the phased array, two horizontally polarized (HPol) antennas are positioned, and four additional vertically polarized antennas are distributed along the power string, extending to a depth of 45 meters. The two additional helper strings sit $\sim$\SI{34.6}{m} horizontal distance from the power string, forming a unilateral triangle. Each helper string has two VPol antennas and an HPol antenna at \SI{144}{m} depth with \SI{1}{m} spacing. Noise is added to the traces using the realistic spectrum shown in Figure~\ref{fig_noise} and scaled to a noise temperature of $T_\text{noise} =$ \SI{300}{K}. Additionally, the neutrino signals are filtered using the same filter shape as the noise spectrum, representing the signal chain response. This choice ensures that the signal and noise are filtered consistently.

We performed an end-to-end simulation of neutrinos with various energies and incoming directions. In the following, we show the reconstruction results for three typical examples with varying event quality, but all fulfill the trigger criteria. The MC true parameter values of the events are shown in Table~\ref{tab_events}. Figure~\ref{fig_signal} shows the simulated neutrino signal in all 16 antennas for Event 3 with one realization of noise. 

\begin{table}[!htb]
    \centering
    \begin{tabular}{|c||c|c|c|c|c|c|c|} 
        \hline
        Event & $E_\text{shower}$ & $\theta_\nu$ & $\phi_\nu$ & $r_\text{vertex}$ & $\theta_\text{vertex}$ & $\phi_\text{vertex}$ & $t_0$ \\ \hline \hline
        1 & $ 15\,\text{PeV} $ & $ 62^{\circ} $ & $ 322^{\circ} $ & $ 220\,\text{m} $ & $ 149^{\circ} $ & $ -70^{\circ} $ & $ 0\,\text{ns} $ \\ \hline
        2 & $ 640\,\text{PeV} $ & $ 64^{\circ} $ & $ 312^{\circ} $ & $ 2576\,\text{m} $ & $ 120^{\circ} $ & $ -57^{\circ} $ & $ 0\,\text{ns} $ \\ \hline
        3 & $ 18\,\text{PeV} $ & $ 91^{\circ} $ & $ 191^{\circ} $ & $ 321\,\text{m} $ & $ 158^{\circ} $ & $ -168^{\circ} $ & $ 0\,\text{ns} $ \\ \hline
    \end{tabular}
    \caption{Monte-Carlo true parameter values of the three representative example events shown in this work.}
    \label{tab_events}
\end{table}

\subsubsection{Likelihood reconstruction} \label{sec_likelihood_reconstruction}

We estimate the shower parameters by minimizing the minus two log-likelihood. For a signal in many antennas, the minus two log-likelihood for each antenna can be summed to give a likelihood for the entire event:
\begin{equation} \label{eq_minus_two_delta_llh}
	\begin{aligned}
		&-2\ln \mathcal{L}(\boldsymbol{\mu}(\boldsymbol{\theta});\boldsymbol{x},\boldsymbol{\Sigma}) \\ & = \sum_{\text{antennas}} \big(\boldsymbol{x}-\boldsymbol{\mu}(\boldsymbol{\theta})\big)^\mathrm{T} \boldsymbol{\Sigma}^{-1}  \big(\boldsymbol{x}-\boldsymbol{\mu}(\boldsymbol{\theta})\big) + const.
	\end{aligned}
\end{equation}
In a reconstruction, the constants can be ignored since they do not depend on the neutrino signal parameters. In our reconstruction, $\boldsymbol{\mu}(\boldsymbol{\theta})$ is the neutrino signal parameterized by the seven parameters, $\boldsymbol{\theta} = (E_\text{shower}, \theta_\nu, \phi_\nu, r_\text{vertex}, \theta_\text{vertex}, \phi_\text{vertex}, t_0)$, described in Sec.~\ref{sec_neutrino_signal_model} and $\boldsymbol{x}$ the simulated signal in the antennas including noise. 

We use the iminuit~\cite{iminuit} Python library, specifically the MIGRAD~\cite{James:1975dr} algorithm, to minimize Eq.~\eqref{eq_minus_two_delta_llh}. The minimizer is initialized at the MC true parameters, assuming that the global optimum of the likelihood can be found. This guarantees fast and stable convergence and allows us to study the impact of the likelihood itself, independent of the rest of the reconstruction chain. Strategies exist for finding the global minimum when the true parameters are not known, e.g., by first only reconstructing the signal arrival direction, which was successfully done in previous work~\cite{Aguilar:2021uzt,Plaisier:2023cxz} or using neural networks to get an initial estimate of the parameters~\cite{Glaser:2022lky}. We leave this to later work as it does not affect the conclusions of this paper. 

The Likelihood not only allows us to reconstruct the neutrino parameters, but also provides a goodness-of-fit measure. If the signal model correctly describes the signal present in the data, the reconstructed minus two log-likelihood, $-2\ln{\mathcal{L}(\boldsymbol{\mu}(\hat{\boldsymbol{\theta}});\boldsymbol{x},\boldsymbol{\Sigma})}$, should follow a chi-square distribution with degrees of freedom equal to the rank of the inverse covariance matrix (or two times the number of non-zero amplitudes in the noise spectrum) minus the number of reconstructed parameters. From this, a p-value for the signal can be calculated, which is crucial for verifying that a reconstructed signal is in agreement with the data. In this simulation study, we verified that all p-values are within the expected distribution.

\begin{figure*}[t]
	\centering
	\includegraphics[width=1\textwidth,trim={0 0.4cm 0 0.3cm},clip]{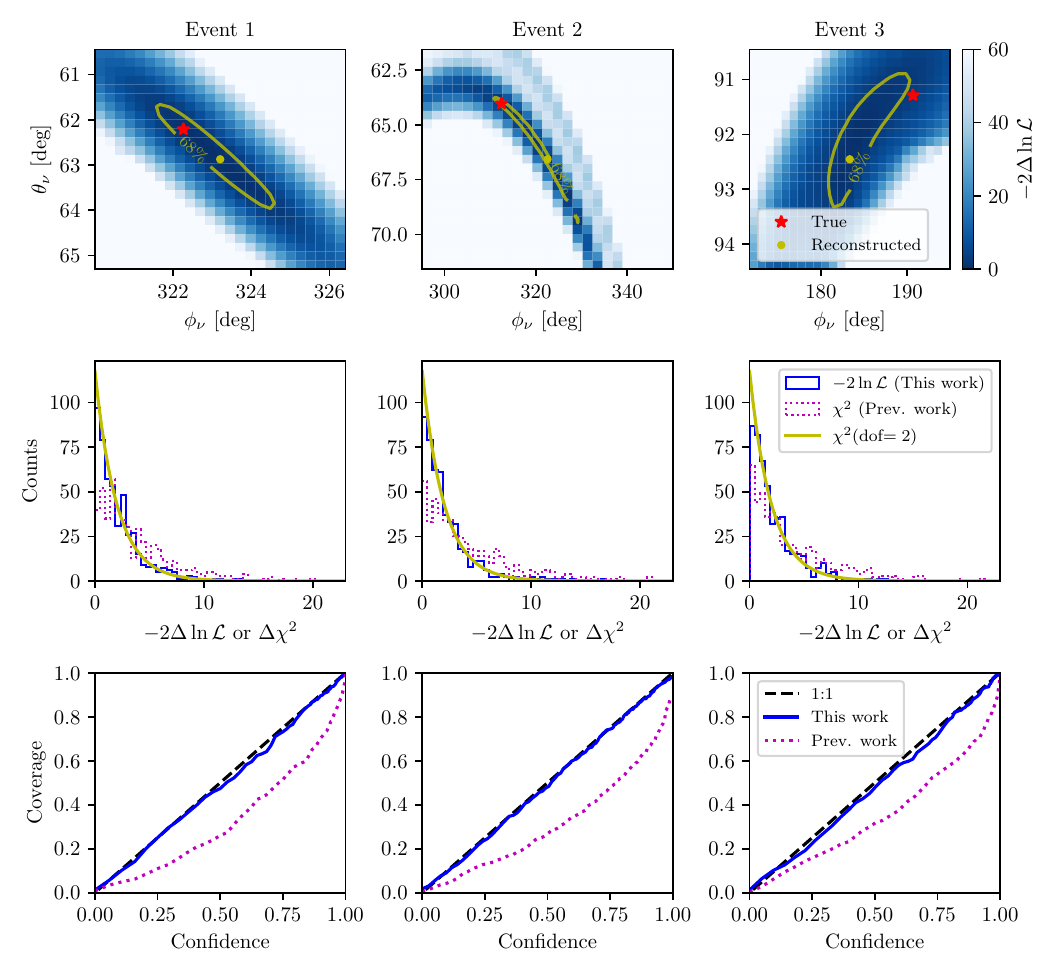}
	\caption{\textit{Top row}: Profile likelihood scans for the three example events in Table~\ref{tab_events} with one noise realization each. \textit{Middle row}: Distribution of $-2\Delta\ln{\mathcal{L}}$ values between the reconstructed neutrino parameters and the true arrival direction profiled over the nuisance parameters for the three events with 500 different noise realizations each. A $\chi^2$-distribution with 2 degrees of freedom is shown for reference. The dotted histograms show the $\Delta\chi^2$ values between the reconstructed and true (profiled over the nuisance parameters)  parameter values using the $\chi^2$ as the objective function (see Sec.~\ref{sec_chi2_comparison}). \textit{Bottom row}: Coverage for any confidence limit for the three events using the $-2\ln{\mathcal{L}}$ or the $\chi^2$ as the objective function in the reconstruction and to estimate the uncertainties.}
	\label{fig_scans_and_coverage}
\end{figure*}

\subsubsection{Profile likelihood scans} \label{sec_profile_llh}

For astrophysical analyses, the reconstructed parameters of interest are often the neutrino arrival direction or the energy. In this case, one can focus on the parameters of interest and treat the rest as nuisance parameters.

To obtain uncertainties on the reconstructed parameters of interest, a profile likelihood scan is performed. The procedure is as follows: First, the reconstruction is performed with all parameters free. This gives the best-fit parameters and a best-fit likelihood value. Second, a scan is performed on a grid in the parameters of interest, either arrival direction or energy. For every point in the grid, a fit is performed with the arrival direction or energy fixed to the grid value and the nuisance parameters free. This gives a likelihood for an alternative hypothesis. The minus two delta log-likelihood can then be used to determine confidence limits (uncertainty contours) for, e.g., $1\sigma$, $2\sigma$, and $3\sigma$ confidence levels ($68.3\%$, $95.5\%$, and $99.7\%$ containment, respectively), using Wilks' theorem. This is shown for the neutrino arrival direction in Fig.~\ref{fig_scans_and_coverage} (Top row) for the three example events. For two degrees of freedom ($\theta_\nu$ and $\phi_\nu$), the $68.3\%$ uncertainty contour corresponds to a  $-2\Delta\ln{\mathcal{L}}$ of $2.30$.

To investigate whether the uncertainties estimated with this method are correct, we quantify their coverage, i.e., the expected number of times the MC true values of the parameters fall within the uncertainties. This is done by repeating the reconstruction of the same event with different realizations of noise 500 times. The best-fit likelihood values are then compared to the likelihoods of the MC true neutrino arrival direction profiled over the nuisance parameters. According to Wilks' theorem, the $-2\Delta\ln{\mathcal{L}}$ should then follow a $\chi^2$ distribution with two degrees of freedom, which we used to convert the likelihood to uncertainty contours. This is shown in Fig.~\ref{fig_scans_and_coverage} (Middle row) and shows good agreement between the expected and real distributions. By comparing the integrated $\chi^2$ distribution to the integrated distribution of $-2\Delta\ln{\mathcal{L}}$ values, we obtain the coverage for any confidence level, which is shown in Fig.~\ref{fig_scans_and_coverage} (Bottom row). The figure shows that the coverage is correct for all confidence levels. Hence, using the likelihood proposed in this paper, we obtain uncertainties on the reconstructed parameters with correct coverage on an event-by-event basis.

\begin{table*} [t]
\begin{center}
\begin{tabular}{|c|c||c|c|c|c|c|c|c|} 
\hline
Event & Method & $\sigma_{E_\nu}$ & $\sigma_{\theta_\nu}$ & $\sigma_{\phi_\nu}$ & $\sigma_{r_\text{vertex}}$ & $\sigma_{\theta_\text{vertex}}$ & $\sigma_{\phi_\text{vertex}}$ & $\sigma_{t_0}$ \\ \hline \hline
1
& \begin{tabular}[c]{@{}c@{}} $\mathcal{L}$ \\  $\chi^2$\end{tabular}
& \begin{tabular}[c]{@{}r@{}} 5.7\% \\ 6.8\% \end{tabular}
& \begin{tabular}[c]{@{}r@{}} 0.8$^{\circ}$ \\ 1.1$^{\circ}$ \end{tabular}
& \begin{tabular}[c]{@{}r@{}} 1.0$^{\circ}$ \\ 1.4$^{\circ}$ \end{tabular}
& \begin{tabular}[c]{@{}r@{}} 4.0\,m \\ 4.5\,m \end{tabular}
& \begin{tabular}[c]{@{}r@{}} 0.769$^{\circ}$ \\ 0.861$^{\circ}$ \end{tabular}
& \begin{tabular}[c]{@{}r@{}} 0.120$^{\circ}$ \\ 0.138$^{\circ}$ \end{tabular}
& \begin{tabular}[c]{@{}r@{}} 0.04\,ns \\ 0.05\,ns \end{tabular}
\\ \hline
2
& \begin{tabular}[c]{@{}c@{}} $\mathcal{L}$ \\  $\chi^2$\end{tabular}
& \begin{tabular}[c]{@{}r@{}} 11.8\% \\ 18.3\% \end{tabular}
& \begin{tabular}[c]{@{}r@{}} 1.6$^{\circ}$ \\ 1.7$^{\circ}$ \end{tabular}
& \begin{tabular}[c]{@{}r@{}} 8.8$^{\circ}$ \\ 9.0$^{\circ}$ \end{tabular}
& \begin{tabular}[c]{@{}r@{}} 79.6\,m \\ 131.0\,m \end{tabular}
& \begin{tabular}[c]{@{}r@{}} 0.066$^{\circ}$ \\ 0.107$^{\circ}$ \end{tabular}
& \begin{tabular}[c]{@{}r@{}} 0.024$^{\circ}$ \\ 0.026$^{\circ}$ \end{tabular}
& \begin{tabular}[c]{@{}r@{}} 0.04\,ns \\ 0.05\,ns \end{tabular}
\\ \hline
3
& \begin{tabular}[c]{@{}c@{}} $\mathcal{L}$ \\  $\chi^2$\end{tabular}
& \begin{tabular}[c]{@{}r@{}} 11.6\% \\ 13.1\% \end{tabular}
& \begin{tabular}[c]{@{}r@{}} 0.6$^{\circ}$ \\ 0.8$^{\circ}$ \end{tabular}
& \begin{tabular}[c]{@{}r@{}} 3.7$^{\circ}$ \\ 4.7$^{\circ}$ \end{tabular}
& \begin{tabular}[c]{@{}r@{}} 20.5\,m \\ 23.8\,m \end{tabular}
& \begin{tabular}[c]{@{}r@{}} 0.988$^{\circ}$ \\ 1.141$^{\circ}$ \end{tabular}
& \begin{tabular}[c]{@{}r@{}} 0.129$^{\circ}$ \\ 0.150$^{\circ}$ \end{tabular}
& \begin{tabular}[c]{@{}r@{}} 0.11\,ns \\ 0.12\,ns \end{tabular}
\\ \hline
\end{tabular}
\caption{Standard deviations, i.e., uncertainties, of best-fit signal parameters for the three neutrino events in Table~\ref{tab_events} with 500 different realizations of noise each using the $-2\ln{\mathcal{L}}$ or the $\chi^2$ as the objective function in the reconstruction. The uncertainties on the shower energy is given as a percentage of the true shower energy.}
\label{table_llh_vs_chi2}
\end{center}
\end{table*}

Additionally, the likelihood values of the best-fit parameters were compared to the true parameter likelihood values, without profiling over the nuisance parameters. This again gives $-2\Delta \ln{\mathcal{L}}$ distributions that follow the expected $\chi^2$ distributions (not shown here) with 7 degrees of freedom, and the coverages are correct for all confidence levels. Hence, if uncertainty contours were drawn in the 7-dimensional parameter space, they would be correct.

\subsubsection{Importance of considering bin-to-bin correlations} \label{sec_chi2_comparison}
To verify the importance of using the correct probabilistic description of noise in radio detectors, we perform the same reconstruction ignoring noise correlations, i.e., we use a chi-square as expressed in Eq.~\eqref{eq_chi2} as the objective function in the reconstruction. This has been used as the objective function in all previous work employing the forward folding technique, e.g.,~\cite{Glaser:2019rxw,Arianna:2021lnr,Gaswint:2021smu,Plaisier:2023cxz}. The resulting $\Delta\chi^2$ distribution (true arrival direction profiled over nuisance parameters minus best fit) for 500 trials is shown in the middle row of Fig.~\ref{fig_scans_and_coverage}. The distribution does not follow a $\chi^2$-distribution with 2 degrees of freedom since the objective function does not take into account the correlations in the noise. It can hence not be used to estimate the uncertainty on the reconstructed parameters. In the bottom row of Fig.~\ref{fig_scans_and_coverage}, it is shown that if uncertainties are estimated assuming the $\Delta\chi^2$ is $\chi^2$-distributed, it would result in under-coverage for the three events.

In addition to not being able to estimate uncertainties, the uncertainties themselves are also larger when minimizing a $\chi^2$ compared to a $-2\ln{\mathcal{L}}$. The standard deviations of the 7 reconstructed parameters for the 500 reconstructions of the same neutrino signals with different realizations of noise using both methods are shown in Table~\ref{table_llh_vs_chi2}. It shows that the spread of best-fit parameters is consistently lower when minimizing the $-2\ln{\mathcal{L}}$ compared to the $\chi^2$. The reason for this is that by taking the information about the correlations in the noise into account, the fitted parameters are better constrained close to the true values. A similar study for the biases of the two methods is shown in Appendix~\ref{sec_neutrino_reco_biases}, which shows a similar improvement for most parameters using the likelihood. Hence, the likelihood method proposed in this paper substantially improves the reconstruction uncertainties, with the magnitude of the improvement depending on the frequency spectra of the noise and signals.

\subsection{Fast uncertainty estimation} \label{sec_fisher_information_matrix}

In the following, we discuss how the noise model and likelihood description enable us to estimate the reconstruction resolution of radio experiments in a computationally efficient way, which can be used for the optimization of future detectors.

Estimating the reconstruction capabilities of an experiment can be done in several ways. The most robust, though computationally expensive, method is to generate a large Monte-Carlo dataset and reconstruct the events to determine the uncertainties on the reconstructed parameters. This method has successfully been used in the past, for example, to estimate the reconstruction resolution of in-ice neutrino detectors~\cite{Glaser:2022lky, Plaisier:2023cxz, Aguilar:2021uzt}. It is, however, often extremely computationally expensive, especially when averaging over a large sample of events, and hence too inconvenient for, e.g., detector optimization, where many different detector layouts are investigated.

\begin{figure*}[t]
	\centering
	\includegraphics[width=1\textwidth,trim={0 0.4cm 0 0.3cm},clip]{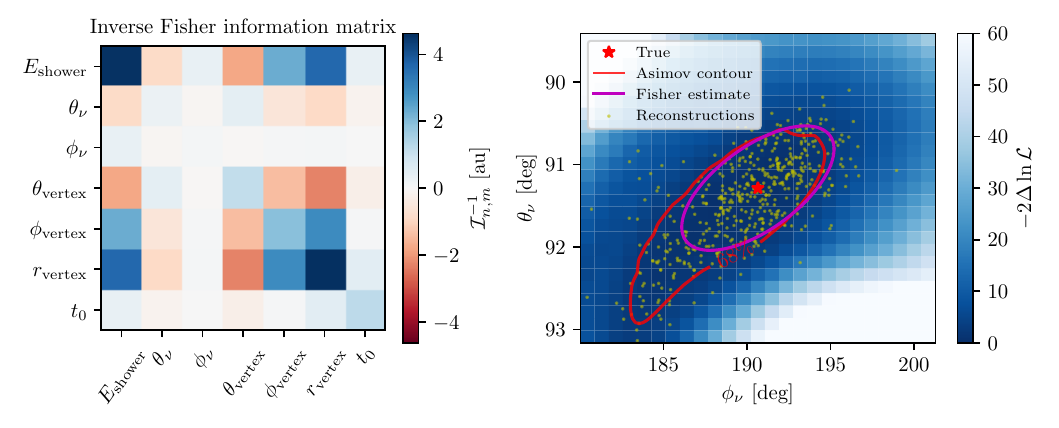}
	\caption{\textit{Left}: Inverse Fisher information matrix for Event 3 in Table~\ref{tab_events}. The units of the entries are the product of the units of the two components. Here the units have been scaled for visibility and are \SI{1}{PeV}, \SI{1}{\degree}, \SI{10}{\degree}, \SI{1}{\degree}, \SI{0.1}{\degree}, \SI{10}{m}, and \SI{0.1}{ns} for the seven parameters, respectively. \textit{Right}: Asimov profile likelihood scan for the same event with $68\%$ containment contour and 500 reconstructions with different realizations of noise. The marginalized Fisher information estimate of the uncertainty ($68\%$ containment) on the neutrino arrival direction is shown centered around the MC true parameter values.}
	\label{fig_fisher_information_matrix}
\end{figure*}

The likelihood description presented in this work enables two additional methods. The first of which is a so-called \emph{Asimov} likelihood scan~\cite{Cowan:2010js}. The Asimov dataset is the `most representative' dataset, i.e., the average of all traces, for a given set of true parameter values. It is, hence, simply equal to the signal prediction in an antenna without any noise added. We can then perform a likelihood scan, or a profile likelihood scan as described in Sec.~\ref{sec_profile_llh}, treating the Monte-Carlo truth signal as the observed trace. The resulting contour is the median reconstruction uncertainty for the given set of true parameter values and should match the distribution of reconstructed values if we were to do many reconstructions of the same event with different realizations of noise. The Asimov profile likelihood contour for Event 3 in Table~\ref{tab_events} is shown in Fig.~\ref{fig_fisher_information_matrix} (Right) along with 500 reconstructions with different realizations of noise. The figure shows that the Asimov contour captures the shape of the distribution well. 

The most computationally efficient way to estimate reconstruction uncertainty, enabled by the likelihood description, is to calculate the Fisher information. The Fisher information matrix tells us how much information a sample from a distribution, e.g., a deterministic signal plus random noise, carries about the parameters that model the distribution, e.g., the signal parameters. For data described by a multivariate normal distribution, where the covariance matrix does not depend on the signal parameters, i.e., signals in radio detectors, the Fisher information matrix is given by
\begin{equation}
	\begin{aligned}
		\mathcal{I}_{nm} &= \sum_{\text{antennas}} \frac{\partial \boldsymbol{\mu}^\mathrm{T}}{\partial \theta_n} \boldsymbol{\Sigma}^{-1}  \frac{\partial \boldsymbol{\mu}}{\partial \theta_m},
	\end{aligned}
\end{equation}
which depends only on the derivatives of the signal with respect to the parameters and the covariance matrix of the noise, and it can easily be summed over the antennas measuring the signal. The Crámer-Rao bound~\cite{a61aa5fe-d74a-3133-bed2-f35c3c555015, MR15748} states that for an unbiased estimator, the inverse Fisher information matrix is a lower bound on the covariance matrix of the parameters:
\begin{equation}
	\begin{aligned}
		\text{Cov}(\boldsymbol{\theta}) \geq \mathcal{I}^{-1}.
	\end{aligned}
\end{equation}
This bound is saturated for large signal amplitudes. A maximum likelihood estimator is, however, not guaranteed to be unbiased, which can be seen from the asymmetric distribution of reconstructed parameters in Fig.~\ref{fig_fisher_information_matrix} (Right) and quantized in Appendix~\ref{sec_neutrino_reco_biases}. In this case, the Crámer-Rao bound can be slightly broken, and the inequality turns into an approximate estimate of the covariance matrix. We note that this is of little concern for optimization studies because by optimizing the estimated uncertainties, the real uncertainties should also improve, which can be verified with a full sensitivity study of an initial and optimized detector design.

The inverse Fisher information matrix for Event 3 in Table~\ref{tab_events} has been calculated using numerical differentiation and is shown in Fig.~\ref{fig_fisher_information_matrix} (Left). By selecting the $2\times2$ sub-matrix associated with the neutrino arrival direction, we obtain the marginalized covariance matrix which parameterizes an ellipse corresponding to the uncertainties of these parameters. The Fisher estimate of the uncertainties is shown in Fig.~\ref{fig_fisher_information_matrix} (Right). We see that the Fisher estimate is an elliptical (2nd order) approximation that does not capture the full asymmetries of the distribution, but its area sufficiently matches the Asimov contour. Since the Fisher estimate can be calculated many orders of magnitude faster than the full reconstruction and profile likelihood scan, it is useful in analyses where the exact shapes of the uncertainties are superfluous and only their approximate magnitudes are of interest.

The Fisher information matrix is an important tool in the prospects of end-to-end detector optimization of radio detectors~\cite{Glaser:2023udy}. With the use of differentiable programming, e.g., PyTorch or JAX, and deep learning-based surrogate models, a fully differentiable radio neutrino signal pipeline can be implemented. The signal will then be automatically differentiable with respect to both the signal parameters and the antenna positions and orientations. Additionally, the Fisher uncertainty calculation can be implemented as a differential proxy for the reconstruction uncertainty, which means that the gradients of the reconstruction uncertainty with respect to the detector layout can be calculated using backpropagation. It hence allows us to efficiently optimize the layout of radio detectors by minimizing the reconstruction uncertainty with respect to the detector parameters. This concept holds great potential for improving the scientific outcome of future radio detectors and will be the subject of future studies.

\subsection{Electric-field reconstruction for cosmic rays} \label{sec_electric_field_reconstruction}
The likelihood can improve another key reconstruction task. Air-shower detectors are typically built with dual-polarized antennas, i.e., two antennas in one mechanical structure that measure two orthogonal polarization components of the electric field simultaneously~\cite{Huege:2016veh}. This measurement, together with the knowledge of the arrival direction, allows for the reconstruction of the incident electric field. The electric-field reconstruction constitutes the low-level component of a full cosmic-ray air shower reconstruction, where, e.g., the energy fluences calculated from the electric field at each antenna are combined to estimate the air-shower energy (e.g.~\cite{PierreAuger:2015hbf}) and $X_\text{max}$ (e.g.~\cite{Buitink:2014eqa}) of the shower. Hence, an improvement in this first step will result in an improvement in the reconstruction of the key observables, the shower energy and $X_\text{max}$. 

In this section, we compare the likelihood method to the currently most widely used method, which obtains the electric field via unfolding (see Sec.~\ref{sec:noise_subtraction} for details). This method does not utilize the full time-domain information of the waveform, which limits reconstruction resolution, and it is known to be biased at low SNR. We also compare our method to the \emph{forward folding} method proposed in Ref~\cite{Glaser:2019rxw}, which uses an analytic electric field model and minimizes a $\chi^2$ objective function without considering correlations in the noise. Our method adopts the same analytic electric field model, but minimizes the $-2\ln{\mathcal{L}}$ instead, hence taking noise correlations into account. In the following, we show that this improves the polarization and energy fluence reconstruction uncertainties over both previous methods in a simulation study.

The results of this section have been previously shown at a conference~\cite{Glaser:2025Ey}.

\subsubsection{Simulation dataset}

To evaluate the performance of the reconstruction methods discussed here, we generate a simulation dataset of cosmic-ray air showers using the CoREAS extension~\cite{Huege:2013vt} of the CORSIKA framework. The dataset consists of a large set of proton-initiated showers with energies uniformly distributed in $\log_{10}(E)$ between $10^{15}$ and $10^{19}$ \text{eV}, and arrival directions uniformly distributed over a zenith angle range of \SI{60}{\degree} to \SI{85}{\degree}. We simulate the showers for the magnetic field and atmospheric conditions at the South Pole. For each air shower, these simulations produce the time-dependent electric field at 160 observer positions placed on the ground in the standard star-shaped grid around the shower axis. We use these electric fields directly in the subsequent calculations and do not apply any additional randomization of positions and interpolation. As we present all our findings as a function of signal-to-noise ratio, this is the most accurate approach. 

Each simulated electric field is observed by a dual-polarized antenna. The antenna is a logarithmic periodic dipole antenna (LPDA) that measures with a sampling rate of \SI{500}{MHz} in the frequency band from \SI{30}{MHz} to \SI{80}{MHz} with two readout channels, one sensitive to north-south polarized electric fields and one sensitive to east-west polarizations. We employ a simple analytical approximation of the LPDA response available in NuRadioReco~\cite{Glaser:2019rxw}, which we use to simulate the response of the antenna to the electric field.

\begin{figure}[b]
    \centering
    \includegraphics[width=1\linewidth]{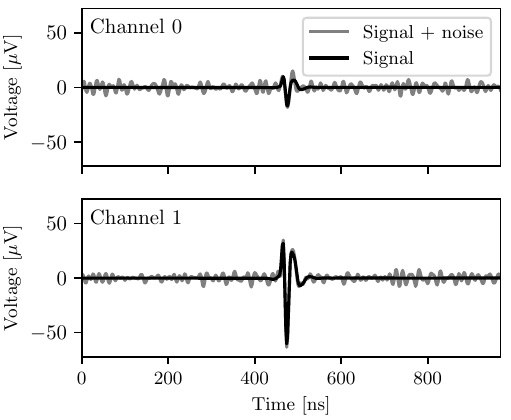}
    \caption{Example air-shower signal from CoREAS simulation as measured by a dual-polarized antenna. The black curve shows the noiseless signal, whereas the gray curve shows the signal with measurement noise.}
    \label{fig:efield_data}
\end{figure}

Finally, we add noise to the traces and apply several filters to simulate hardware- and analysis-level filters, which remove environmental noise at low and high frequencies. The generated noise follows a smooth spectrum in the \SI{30}{MHz} to \SI{80}{MHz} range, implemented using a third-order Butterworth high-pass filter and an eighth-order Butterworth low-pass filter. This frequency band is chosen as it represents a typical range for air-shower radio arrays~\cite{Huege:2023pfb}. Additionally, as very low SNR observers are overrepresented, we randomly skip some of them to make the distribution of events over SNR more uniform. The SNR is calculated using the maximum absolute amplitude of the noiseless traces divided by the standard deviation of the noise. The resulting dataset contains 73,459 simulated cosmic-ray electric-field waveforms with SNR values below 20, which we will reconstruct in the following sections. We put our emphasis on low SNRs, as this is the region that is difficult to reconstruct. An example of the resulting traces as measured by the dual-polarized antenna is shown in Fig.~\ref{fig:efield_data}. 

For the purpose of this study, we assume the cosmic-ray arrival direction is known from external reconstruction (e.g., from triangulation using the signal times in several observers), and the signal arrival time is known to within \SI{30}{ns}, which we call the \textit{search window}. This is a reasonable assumption, as the arrival direction can often be well constrained by fitting a wavefront to the arrival times of the signal in antennas with high SNR. The wavefront can then be extrapolated to estimate the arrival time in antennas with lower SNR.

\subsubsection{Previous method: unfolding and noise subtraction} \label{sec:noise_subtraction}
The \textit{unfolding} and \textit{noise subtraction} method is the most commonly used method to estimate the fluence and polarization of an electric field in a dual-polarized antenna~\cite{Glaser:2017ctn}. The method comprises three simple steps. First, the measured trace is unfolded with the antenna response for the known arrival direction of the shower. This obtains an estimate of the three-dimensional (polarizations), time-dependent electric field, $\boldsymbol{E}$ of shape $n_\text{pol} \times n_t$, that includes unfolded noise from the trace. It is worth noting that the unfolded noise is not a correct physical representation of the radio noise that the antenna sees, since all noise is unfolded as if it is coming from the direction of the cosmic ray and includes noise from the electronics. This can lead to biases in reconstructed quantities.

To estimate the time of the peak of the electric field, a Hilbert envelope is applied in a \textit{search window}, the norm of the three-dimensional envelope is calculated, and the peak is identified as $t_{\text{peak}}$. In this work, the search window is defined as $t_{\text{true}}\pm$\SI{30}{ns} where $t_{\text{true}}$ is the peak of the norm of the Hilbert envelope of the noiseless electric field. A \textit{signal window} is then defined as $[t_1,t_2]$ at $\pm$\SI{30}{ns} around $t_{\text{peak}}$, and a \textit{noise window} is defined as $[t_3,t_4]$, which is a \SI{200}{ns} period outside the signal window. The electric field energy fluence of the three components is then estimated as
\begin{equation} \label{eq_noise_subtraction}
	\begin{aligned}
		\widehat{f}_\text{pol} = \varepsilon_0 c \Bigg( &
            \delta t \sum_{t_1 \le t_n \le t_2} |\boldsymbol{E}_{\text{pol},n}|^2 
            \\ &- \delta t \frac{t_2 - t_1}{t_4 - t_3} 
           \sum_{t_3 \le t_n \le t_4} |\boldsymbol{E}_{\text{pol},n}|^2 
            \Bigg),
	\end{aligned}
\end{equation}
where $\varepsilon_0$ is the dielectric constant and $c$ is the speed of light. The method hence estimates the total fluence in the signal window and subtracts the contribution from the noise estimated from the noise window. We set $f_\text{pol}$ to zero in cases where the estimated noise fluence is larger than the fluence in the signal window, to avoid non-physical negative fluence estimates.

The basis vectors of the coordinate system chosen for three polarizations of the electric field are along the shower axis, $\hat{r}$ (with $\boldsymbol{E}_r$ assumed to be $0$), perpendicular to the shower axis in the vertical plane, $\hat{\theta}$, and perpendicular to the shower axis in the horizontal plane, $\hat{\phi}$. The estimate of the total fluence is then $\widehat{f}_\text{tot} = \widehat{f}_\theta+\widehat{f}_\phi$ and the polarization of the electric field is
\begin{equation}
    \widehat{P} = \arctan\bigg(\sqrt{\widehat{f}_\phi}/\sqrt{\widehat{f}_\theta}\bigg).
\end{equation}
We note that using this convention, all measured polarizations will be between \SI{0}{\degree} and \SI{90}{\degree}.

The uncertainties of the reconstructed fluences are estimated by assuming the noise in the noise window is Gaussian and uncorrelated, and the uncertainty on the polarization is derived using error propagation.

\subsubsection{Forward folding reconstructions: chi-square and likelihood} \label{sec_forward_folding_likelihood}

The benefits of forward folding methods are twofold. First, it avoids unfolding system and ambient noise as if it was coming from one specific direction, and second, it constrains the signal to a parameterized model representing the physical process being observed. The parameterized model can either be a full simulation (as in Section~\ref{sec_neutrino_reconstruction}) or a simpler approximation. Since cosmic-ray air-shower simulations are computationally expensive, we employ a simple frequency domain parameterization of the electric field, which has been shown to describe radio pulses from cosmic-ray air showers well~\cite{Glaser:2019rxw}:
\begin{equation}
\begin{pmatrix} \mathcal{E}_{\theta,k} \\ \mathcal{E}_{\phi,k} \end{pmatrix}
= \begin{pmatrix} A_\theta \\ A_\phi \end{pmatrix} 10^{\alpha \cdot f_k + \beta (f_k - f_\text{offset})^2} e^{-i (2\pi f_k \cdot t_\text{offset} - \psi)},
\end{equation}
where $\mathcal{E}$ is the electric field in the frequency domain, $f_k$ are the frequencies, and $A_\text{pol}$ are the electric field amplitudes. The frequency dependence of $\mathcal{E}$ is controlled by $\alpha$, which sets the slope of the electric field in the frequency domain and is often constrained to negative values, and $\beta$, which controls a second-order correction where $f_\text{offset}$ is fixed to \SI{30}{MHz}. The last exponential controls the complex phases, where $t_\text{offset}$ is the time of the pulse relative to the start of the trace, and $\psi$ is a phase offset. Additionally, we numerically re-normalize the electric field, such that the parameters we reconstruct are directly the fluences of the two polarizations, $f_\theta$ and $f_\phi$. In our reconstruction, we allow $f_\theta$ and $f_\phi$ to take on negative values, corresponding to a sign flip of $A_\theta$ and $A_\phi$, respectively. This makes the method sensitive to polarizations across the \SI{0}{\degree} to \SI{180}{\degree} range (polarization flips of 180 are degenerate with changing $\psi$ by $\pi$). However, for a fair comparison, we restrict it here to the same range as the noise subtraction method in a consistent way. 

For a given set of parameter values, the electric field is multiplied by the vector effective length (VEL) of the antenna for the known arrival direction and filtered accordingly. By applying the inverse Fourier transform, the estimate of the noiseless signal, $\boldsymbol{\mu}$, is obtained, which can then be compared to the measured trace in an objective function. The resulting signal depends on six parameters to be reconstructed, $\boldsymbol{\theta} = (f_\theta, f_\phi, \alpha, \beta, t_\text{offset}, \psi)$.

\begin{figure*}[t]
    \centering
    \includegraphics[width=1\linewidth,trim={0 0.4cm 0 0.3cm},clip]{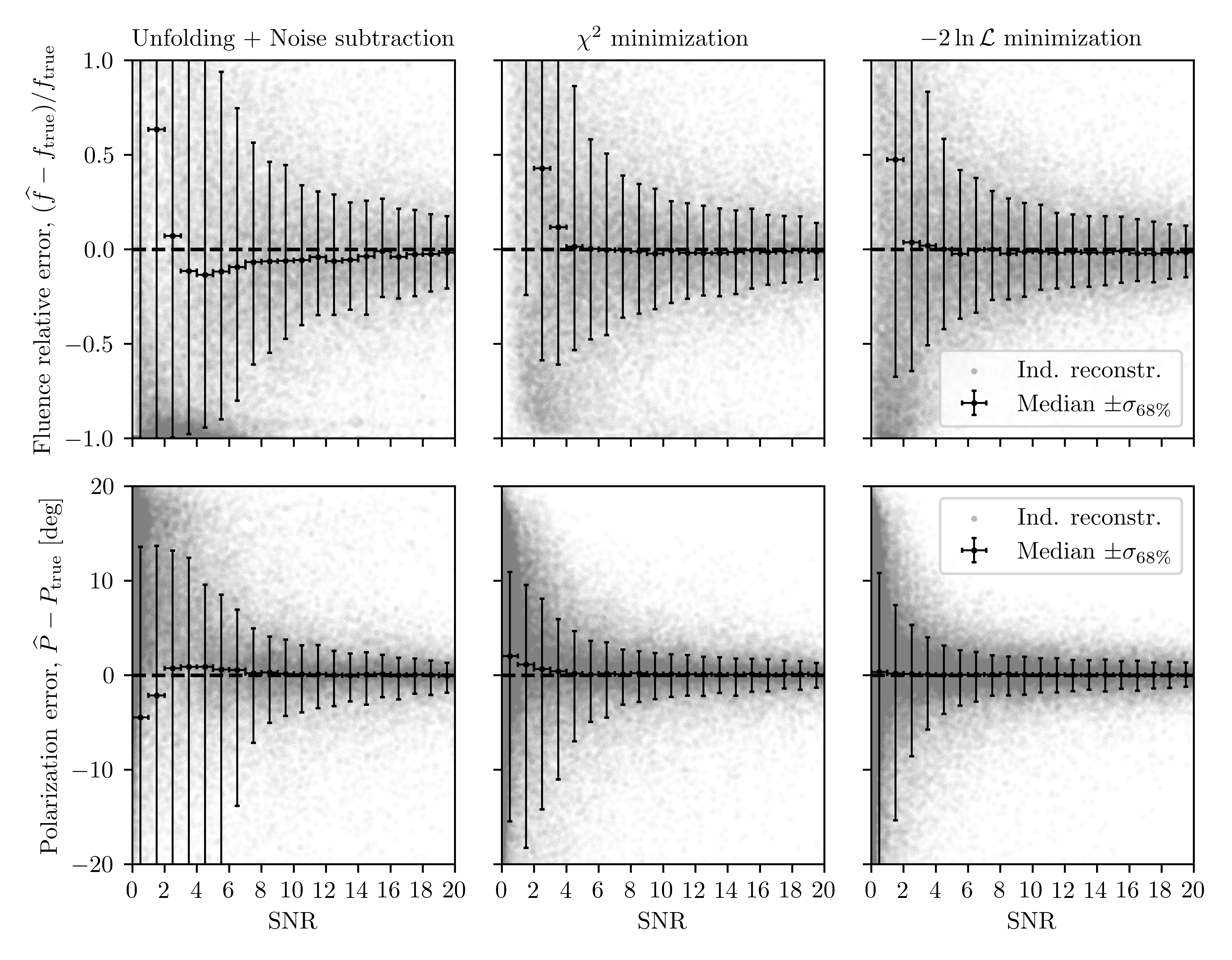}
    \caption{Reconstruction results for the three methods. The relative errors (reconstructed minus true divided by true) of the total fluences and errors (reconstructed minus true) of the polarizations are shown for individual events. Many of the individual reconstructions (transparent gray dots) fall outside the plotted range. The events were binned in SNR to calculate the median, i.e., the bias, and asymmetric central 68\%-quantiles, i.e., the uncertainties ($\pm\sigma_{68\%}$), of each bin.}
    \label{fig:efield_reco_vs_snr}
\end{figure*}

Traditionally, the objective function to be minimized is a $\chi^2$, as expressed in Eq.~\eqref{eq_chi2}, which does not consider the correlation in the noise. We propose minimizing the $-2\ln{\mathcal{L}}$ expressed in Eq.~\eqref{eq_minus_two_delta_llh}, which correctly describes the band-limited nature of the noise. In this work, the minimization algorithms for both objective functions were improved to achieve better stability. For the $-2\ln{\mathcal{L}}$ minimization, the algorithm consists of two minimization processes. In the first, a matched filter (see Appendix~\ref{sec_matched_filter}) is applied in every step of the minimization to analytically profile over amplitudes and efficiently profile over a grid of $t_\text{offset}$ in the search window. This drastically reduces the number of local optima of the likelihood, and allows us to fit the parameters that control the shape of the signal, $\alpha$, $\beta$, $\psi$, and the ratio of the fluences $r_{\theta/\phi}=f_\theta/f_\phi$. In the second minimization, the $-2\ln{\mathcal{L}}$ is minimized with all parameters free, including the time and overall amplitude, to fine-tune the result. Since the $-2\ln{\mathcal{L}}$ has local minima for opposite polarizations, we run the algorithm four times, initialized in each quadrant of $(f_\phi,f_\theta)$ to guarantee convergence.

The $\chi^2$ minimization algorithm was improved in a corresponding way by using a normalized cross-correlation (Eq.~\eqref{eq_cross_correlation}) to profile over time and amplitude in the first minimization process. The normalized cross-correlation is equivalent to a matched filter with a diagonal covariance matrix, i.e., no correlations taken into account. This allows us to evaluate the impact of the objective function directly, i.e., $-2\ln{\mathcal{L}}$ or $\chi^2$, with the same improvements to the minimization algorithm.

For both objective functions, the uncertainties on the reconstructed parameters are extracted from the Hessian matrix at the minimum of the $-2\ln{\mathcal{L}}$ or $\chi^2$. The uncertainty on the polarization can then be estimated using error propagation.

\subsubsection{Reconstruction performances}

\begin{figure*}[t]
    \centering
    \includegraphics[width=1\linewidth,trim={0 0.4cm 0 0.3cm},clip]{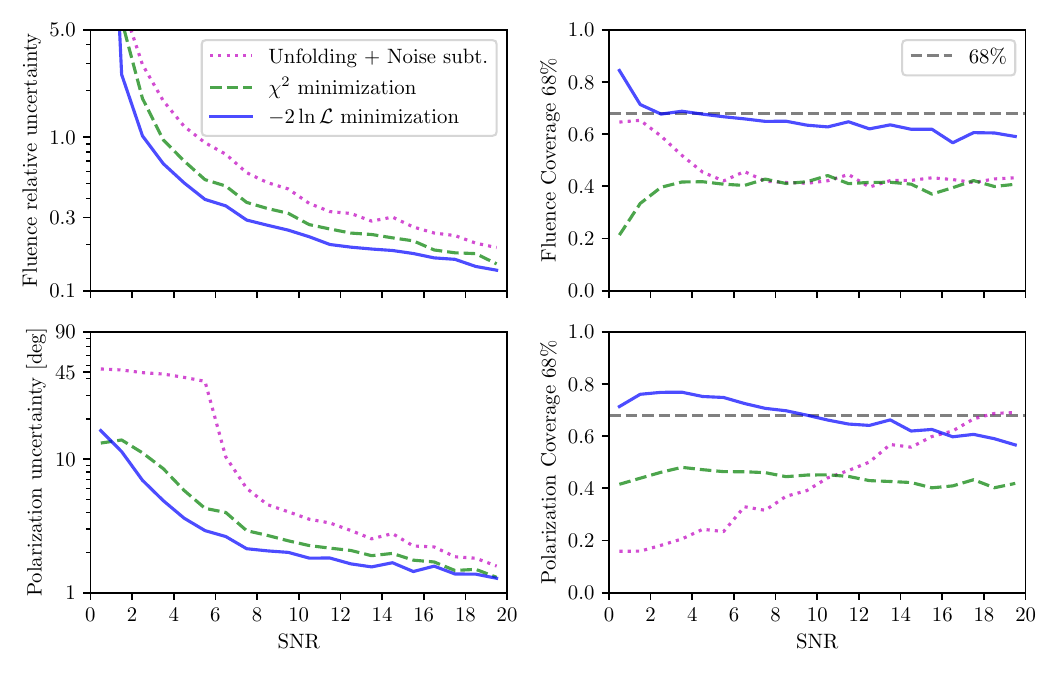}
    \caption{\textit{Left}: Half the width of the asymmetric central 68\%-quantiles, i.e., uncertainties, of the fluence relative uncertainties (reconstructed minus true divided by true) and polarization uncertainties (reconstructed minus true) of the three methods. \textit{Right}: Coverage of the event-by-event estimated 68\% uncertainties of the fluence and polarization for the three methods, i.e., how often the true value is within the estimate plus/minus the estimated uncertainty, which should ideally be 68\%.}
    \label{fig_efield_reco_comparison}
\end{figure*}

All events in the simulation dataset were reconstructed using the three methods discussed in the previous sections. The relative errors (reconstructed minus true divided by true) of the reconstructed fluences and the errors (reconstructed minus true) of the reconstructed polarizations are shown in Fig.~\ref{fig:efield_reco_vs_snr}. The results are binned in SNR, and the medians and central 68\%-quantiles, $\pm\sigma_{68\%}$, of each bin are shown in the figure. The noise subtraction method has two competing biases in the fluence at low SNR. At $\mathrm{SNR}<3$, there is a positive bias originating from the peak-finding, which will always find the noisiest part of the unfolded electric field in the search window when almost no signal is present. The negative bias at $\mathrm{SNR}>3$ is caused by the fluence estimator not taking into account that the signal and noise can be destructively aligned. Additionally, since $f_\theta$ and $f_\phi$ are not allowed to be negative, the reconstructed fluences pile up at 0, and the polarizations pile up at \SI{0}{\degree} and \SI{90}{\degree}, which causes very large errors at low SNR.

The $\chi^2$ minimization method reduces the bias at low SNR for both the fluence and polarization reconstruction. The $-2\ln{\mathcal{L}}$ minimization further improves the bias and remains unbiased in the fluence down to $\mathrm{SNR}>2$, and is effectively unbiased in the polarization. The remaining positive bias in the fluence at very low SNR is caused by the minimization fitting the pulse to the largest fluctuation in the noise when no signal is present.

In order to compare the uncertainties of the three methods, we plot half the width of the central 68\%-quantiles of the errors, which we here call the uncertainties, of each bin in Fig.~\ref{fig_efield_reco_comparison} (Left). Using the likelihood method, the fluence relative uncertainties are decreased by a factor ranging from $1.6$ to $2.3$ in the $5<\mathrm{SNR}<15$ range compared to the noise subtraction method, and the polarization uncertainties are decreased by a factor ranging from $1.7$ to $13.1$ in the same range. Compared to the $\chi^2$ method, the likelihood method decreases fluence relative uncertainties by a factor ranging between $1.2$ to $1.3$ and the polarization uncertainties by a factor ranging between $1.2$ to $1.5$ in the same SNR range. The latter improvement arises purely from changing the objective function, which underlines the importance of taking bin-to-bin correlations into account. The improvement in reconstruction resolution is hence substantial, especially at low SNR.

An estimate of the uncertainty on the reconstructed parameters can be obtained event-by-event from the three methods. To verify if the estimated 68\% uncertainties are correct, we calculate the coverage and show it in Fig.~\ref{fig_efield_reco_comparison} (Right). The coverage is calculated by counting how often the true value is within the estimate plus/minus the estimated 68\% uncertainty, which should ideally be 68\% of the time. Both the noise subtraction and $\chi^2$ methods severely under-predict the uncertainties, whereas the likelihood method predicts the uncertainties well across most SNRs. The remaining slight under-coverage is likely caused by the simplicity of the electric field model, which can not perfectly model the cosmic-ray signals at high SNR.

The unfolding method can be improved by estimating the energy fluence from the noisy electric fields based on Rice distributions as recently proposed in Ref.~\cite{Martinelli:2024bzg}. This method takes into account the band-limited nature of the noise but does not constrain the electric field to a physical model, and keeps the problem of the unfolding of amplifying the noise at small SNR. When applied to our simulation dataset, we observed an equivalent bias and similar uncertainties to the noise subtraction method at low SNR. However, the method correctly estimated the uncertainties across all SNRs, and it may hence be beneficial to use it at high SNRs.

The likelihood method for reconstructing electric fields from several antennas with close proximity to each other is implemented in NuRadioReco. Our method has already been applied to RNO-G cosmic-ray data with high signal-to-noise ratios and gives consistent results to the simpler $\chi^2$ method \cite{Agarwal:2025iyo}, thus showing the applicability of our method to experimental data.

\subsection{Signal identification from background} \label{sec_cosmic_ray_identification}
The final application of the likelihood we present is identification of signals from background. The ratio of the likelihoods for two different hypotheses is the most powerful way of comparing two hypotheses~\cite{NeymanPearson}. Thus, using the likelihood is the most effective method for separating a signal from a background when searching for known signals in radio antenna traces. Separating the signal from the background is a common challenge in all radio detection experiments, as the signal purity is extremely low. The first primary analysis step is to identify the cosmic-ray or neutrino-induced radio signals from background events that are several orders of magnitude more numerous. In the following, we demonstrate the method by conducting a toy study of searching for cosmic-ray signals in in-ice radio neutrino detectors, as one of many possible applications. 

Cosmic-ray signals in radio neutrino detectors make up a substantial background. Since cosmic-ray air showers are several orders of magnitude more abundant than UHE neutrinos, and radio emission from air showers is a well-understood phenomenon, they can effectively be used as a calibration source and end-to-end detector and analysis verification and demonstration. Searches for cosmic rays have been carried out with ARIANNA~\cite{Arianna:2021lnr} using shallowly deployed LPDA antennas, demonstrating the reconstruction capabilities of the experiment. A similar analysis was recently published with RNO-G~\cite{Agarwal:2025iyo, Henrichs:2023XE, Nelles:2024c9}.

When identifying cosmic ray signals, the main background is thermal noise that triggers the detector. The measured event rate of cosmic rays by ARIANNA is on the order of 1 per station per day, however, the stations are often configured to trigger and save traces at a \SI{1}{Hz} rate. The vast majority of triggered events are thus thermal noise fluctuations, and a background rejection rate of $\mathcal{O}(\sim 10^5)$ per signal event is needed. To separate signal from background, previous analyses have relied on correlation scores of known signal templates with the measured trace in one antenna,
\begin{equation} \label{eq_cross_correlation}
    \rho = \max \left( \rho(\Delta n) \right) = \max \left( 
    \frac{\sum_{i}^{n_t} \mu_i \cdot x_{i+\Delta n}}{
    \sqrt{\sum_{i}^{n_t} \mu_i^2} \cdot \sqrt{\sum_{j=\Delta n}^{n_t+\Delta n} x_j^2}}
    \right),
\end{equation}
where $x$ is the measured trace, $\mu$ is the template, and $\Delta n$ represents a time offset. This correlation acts as a test statistic for which a cut is set to determine which triggered events are likely cosmic-ray signals. In the following, we demonstrate, using a toy-model study, that utilizing the maximized two delta log-likelihood between no signal (zeros) and the signal template achieves a better separation between the signal and background.

\begin{figure}[b]
	\centering
	\includegraphics[width=1\linewidth,]{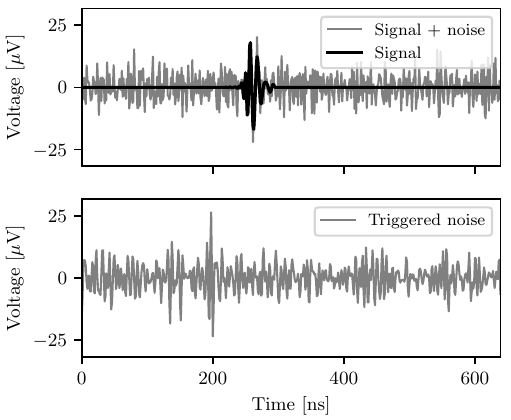}
	\caption{\textit{Top}: Trace with injected cosmic ray signal template and one realization of thermal noise. \textit{Bottom}: Simulated thermal noise that would trigger the detector (\SI{1}{Hz} trigger), which is the main background for identifying cosmic-ray signals in single antennas.}
	\label{fig_CR_trace}
\end{figure}

\begin{figure*}[t]
	\centering
	\includegraphics[width=1\textwidth]{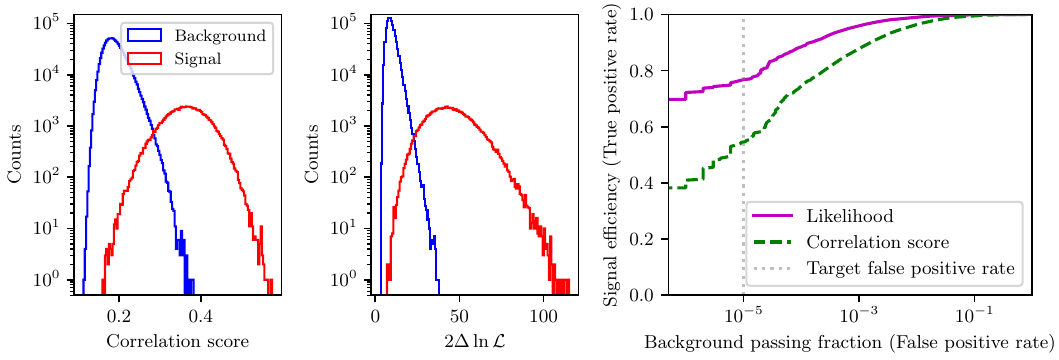}
	\caption{\textit{Left and Middle}: Correlation score and two delta log-likelihood distributions for 1,000,000 background traces containing triggered thermal noise and 100,000 signal traces containing thermal noise and an injected cosmic ray signal. \textit{Right}: Signal efficiency versus background passing fraction (ROC curves) of the toy-model cosmic ray identification analysis using the correlation score and the likelihood as the test statistic. The gray dotted line indicates the target background passing fraction of $10^{-5}$, which gives a 55\% signal efficiency for the correlation score and 77\% for the likelihood.}
	\label{fig_CR_histograms_and_ROC_curves}
\end{figure*}

The toy-model study is carried out as follows: First, a set of 1,000,000 background traces for one antenna is simulated, containing thermal noise that fulfills the trigger criteria. We use a sampling rate of \SI{0.8}{GHz} and a trace length of $n_t=512$ samples, and noise with a temperature of 300 K that follows a smooth spectrum using Butterworth filters at 80 MHz (second-order high-pass) and 220 MHz (tenth-order low-pass). The narrower frequency band is chosen to optimize for cosmic-ray signals, which have more power at lower frequencies than neutrino signals, and targets the frequency range where the LPDAs are most sensitive. We set the trigger criteria such that the station has a \SI{1}{Hz} trigger rate and corresponds to a high-low trigger threshold of 3.3~$\sigma$. Second, we generate a set of 100,000 signal traces, by generating unbiased thermal noise (that typically does not satisfy the trigger criteria) and injecting a cosmic-ray signal with an SNR of 3.3, i.e., at the trigger threshold. An example of a trace with the injected cosmic-ray signal is shown in Fig.~\ref{fig_CR_trace} (Top), and Fig.~\ref{fig_CR_trace} (Bottom) shows a trace of triggered thermal noise with high signalness.

To quantify the performance of identifying signals using the correlation score, we calculate the correlation score for the background and signal traces using the injected signal as the template. Similarly, we calculate the two delta log-likelihood between no signal and the template maximized over time and amplitude. The maximization is performed using the matched filter search approach described in Appendix~\ref{sec_matched_filter}, due to its efficiency for a pre-calculated template. The resulting test statistic distributions are shown in Fig.~\ref{fig_CR_histograms_and_ROC_curves} (left and middle). The histograms show a clear separation between the two classes using both methods, with a smaller overlap for the likelihood method.

To fully assess the gain of the likelihood approach, we calculate the signal efficiency (true positive rate) and a background passing fraction (false positive rate) for any cut on the test statistic. The resulting ROC curves are shown in Fig.~\ref{fig_CR_histograms_and_ROC_curves} (Right). Especially at low background passing fractions, the likelihood achieves a better signal efficiency. At the target background passing fraction of $10^{-5}$, the remaining fraction of cosmic ray events after the cut is 55\% using the correlation score, whereas it is 77\% using the likelihood. For smaller passing fractions, as required for the less abundant neutrinos, the benefit of using the likelihood further increases.

This simple toy-model study demonstrates that the identification of signals in radio antennas can be significantly improved by exploiting the information about the correlations in the noise. The method can be applied in any scenario where known signals are searched for in one or more radio antennas with band-limited noise.

\section{Conclusion}

In this paper, we have proposed a probabilistic description of noise in radio detectors that captures its band-limited/correlated nature. This framework enables the formulation of a likelihood for deterministic signals in such detectors, providing an ideal objective function for reconstruction in cases where a full signal model, or a reasonable proxy, can be formulated. The central code for calculating and optimizing this likelihood with respect to a signal is available in NuRadioMC. Through simulation studies, we have demonstrated that the method accurately reconstructs neutrino signals in in-ice radio detectors as well as electric fields from cosmic-ray air showers using dual-polarized antennas. In both cases, the method provides reliable uncertainty estimates on the reconstructed parameters and substantially improves the reconstruction precision compared to previous approaches. We further highlighted the potential of the likelihood to improve signal efficiency in event selections. Finally, the likelihood formulation enables a fast estimation of reconstruction uncertainties using the Fisher information, enabling differentiable end-to-end detector optimization. The approach presented in this paper has the potential to advance analyses across the field of radio detection of particle physics signals and may play a central role in optimizing future detectors, future discoveries, or precision measurements.

\begin{acknowledgments}
We thank Nils Heyer for generating the neutrino simulation dataset used in Sec~\ref{sec_neutrino_reconstruction} and for useful discussions. We thank the RNO-G collaboration for useful feedback and discussions. The work is supported by the European Union (ERC, NuRadioOpt, 101116890) and by the Swedish Research Council (VR) via the project 2021-05449.
\end{acknowledgments}

\appendix

\section{Frequency domain formulation of multivariate normal distribution} \label{sec_frequency_domain}
It is often convenient to perform calculations in the frequency domain instead of the time domain. We here present a formulation of the noise model (multivariate normal distribution, Eq.~\eqref{eq_multivariate_normal}) using Fourier transforms.

We first define the noise power spectral density as $S_n(\omega_k) \equiv 2 A(\omega_k)^2 \delta f$. From Eq.~\eqref{eq_analytical_inverse}, the following relation can then be derived:
\begin{equation}
        \boldsymbol{a}^\mathrm{T} \boldsymbol{\Sigma}^{-1} \boldsymbol{b} = 4 \mathfrak{Re} \Bigg\{\sum_{k=1}^{n_t/2-1} \frac{\tilde{b}(\omega_k) \, \tilde{a}^*(\omega_k)}{S_n(\omega_k)} \, \delta f \Bigg\},
\end{equation}
where $\tilde{a}(\omega_k)$ and $\tilde{b}(\omega_k)$ are the discrete Fourier transforms (as defined in Eq.~\eqref{eq_rfft}) of real-valued vectors $\boldsymbol{a}$ and  $\boldsymbol{b}$. This relation states that the multiplication of any two real-valued vectors with the inverse of the covariance matrix can be expressed using the Fourier transforms of the vectors and the noise power spectral density. The matrix multiplication in Eq.~\eqref{eq_multivariate_normal} can hence be expressed as
\begin{equation}
\begin{aligned}
    \big(\boldsymbol{x}-\boldsymbol{\mu}(\boldsymbol{\theta})\big)^\mathrm{T} & \boldsymbol{\Sigma}^{-1}   \big(\boldsymbol{x}-\boldsymbol{\mu}(\boldsymbol{\theta})\big)  \\ & = 4 \sum_{k=1}^{n_t/2-1} \frac{|\tilde{x}(\omega_k)-\tilde{\mu}(\omega_k,\boldsymbol{\theta})\big|^2}{S_n(\omega_k)} \, \delta f.
\end{aligned}
\end{equation}
By comparing the definition of $S_n(\omega_k)$ to the eigenvalues of the covariance matrix (e.g., Eq.~\eqref{eq_analytical_covariance_2}), we see that they can be expressed as $\lambda_k = \frac{1}{4 \delta t} S_n(\omega_k)$ with two degenerate eigenvalues per $\omega_k$. The determinant is then
\begin{equation}
    	|{\boldsymbol{\Sigma}}| = \Bigg( \prod_{k=1}^{n_t/2-1} \frac{1}{4 \delta t} S_n(\omega_k) \Bigg)^2.
\end{equation}
The multivariate normal distribution can hence be expressed as
\begin{equation} \label{eq_multivariate_normal_frequency_domain}
\begin{aligned}
    p\big(\boldsymbol{x}; \boldsymbol{\mu}(\boldsymbol{\theta}), S_n \big) = \frac{1}{\sqrt{(2\pi)^{n_t} } \Big( \prod_{k=1}^{n_t/2-1} \frac{1}{4 \delta t} S_n(\omega_k) \Big)} &
    \\ \times \exp \bigg(-\frac{1}{2} 4 \sum_{k=1}^{n_t/2-1} \frac{|\tilde{x}(\omega_k)-\tilde{\mu}(\omega_k,\boldsymbol{\theta})\big|^2}{S_n(\omega_k)} \, \delta f \bigg) &,
\end{aligned}
\end{equation}
from which the likelihood can be derived as in Eq.~\eqref{eq_llh}. This expression is convenient since it circumvents the calculation of the covariance matrix and its inverse. It is also numerically more robust compared to the inversion of the covariance matrix, which has a combination of very large and small numbers. The combination of Fourier transforms and sums is also computationally faster by about a factor of 4 compared to the time-domain calculation.

Similar to the discussion in Sec.~\ref{sec_pseudo_inverse}, if any frequency amplitudes are equal to zero, Eq.~\eqref{eq_multivariate_normal_frequency_domain} is undefined. In this case, the multivariate normal probability can be calculated by letting the sum and product run over all non-zero values of the noise power spectral density, i.e., $S_n(\omega_k)>0$. This is equivalent to using the pseudoinverse of the covariance matrix as in Eq.~\eqref{eq_degenerate_normal}.

\section{Relation to matched filtering} \label{sec_matched_filter}
A common tool used to search for signals in noisy data is called a \textit{matched filter}. It is the filter that maximizes the signal-to-noise ratio for a known signal template in a data stream with an arbitrary known noise spectrum. This method is for instance used in gravitational wave detection, where a large signal template bank is correlated using the \textit{matched filter} with the continuous data stream of the gravitational wave detector, e.g., LIGO, VIRGO, and KAGRA~\cite{Allen:2005fk,Biwer:2018osg,Cannon,Aubin:2020goo,Chu:2020pjv}. In many ways, gravitational wave detection is similar to radio detection of cosmic rays or neutrinos, i.e., searching for rare known pulsed signals in a continuous stream of data with a known noise spectrum.

The \textit{matched filter} can be derived by maximizing the likelihood with respect to the amplitude of a signal template. Let us assume that a signal can be expressed as $\boldsymbol{\mu}(s_0,t_{\rm{s}}, \boldsymbol{\theta}) = s_0 \boldsymbol{\mu_0}(t_{\rm{s}},\boldsymbol{\theta})$, where $s_0$ governs the amplitude of the signal and $\boldsymbol{\mu_0}(t_{\rm{s}},\boldsymbol{\theta})$ is a signal template with normalized amplitude which depends on parameters $\boldsymbol{\theta}$, and $t_{\rm{s}}$ represents a unitary translation/shift of the signal in time. The Fourier transform of the signal template can then be expressed as $\tilde{\mu}(\omega_k; s_0, t_{\rm{s}}, \boldsymbol{\theta}) = s_0 \tilde{\mu}_{00}(\omega_k; \boldsymbol{\theta}) e^{i \omega_k t_{\rm{s}}}$, where $\tilde{\mu}_{00}(\omega_k; \boldsymbol{\theta})$ is the Fourier transform of the normalized template evaluated at $t_{\rm{s}}=0$. We first define the matched filtering quantities:
\begin{equation}
	\begin{aligned}
		y_\text{mf}(t_{\rm{s}}, \boldsymbol{\theta}) & \equiv \boldsymbol{\mu_0}(t_{\rm{s}},\boldsymbol{\theta})^\mathrm{T} \boldsymbol{\Sigma}^{-1} \boldsymbol{x}
        \\ &= \, 4 \mathfrak{Re} \Bigg\{\sum_{k=1}^{n_t/2-1} \frac{\tilde{x}(\omega_k) \tilde{\mu}_{00}(\omega_k;\boldsymbol{\theta})^*}{S_n(\omega_k)} e^{i \omega_k t_{\rm{s}}} \, \delta f \Bigg\},
        \\ y_{\boldsymbol{\mu}}(\boldsymbol{\theta}) & \equiv \boldsymbol{\mu_0}(t_{\rm{s}},\boldsymbol{\theta})^\mathrm{T} \boldsymbol{\Sigma}^{-1} \boldsymbol{\mu_0}(t_{\rm{s}},\boldsymbol{\theta})
        \\ &= 4 \sum_{k=1}^{n_t/2-1} \frac{|\tilde{\mu}_{00}(\omega_k;\boldsymbol{\theta})|^2}{S_n(\omega_k)} \, \delta f,
        \\ y_{\boldsymbol{x}} & \equiv \boldsymbol{x}^\mathrm{T} \boldsymbol{\Sigma}^{-1} \boldsymbol{x}
        \\ &= 4 \sum_{k=1}^{n_t/2-1} \frac{|\tilde{x}(\omega_k)|^2}{S_n(\omega_k)} \, \delta f ,
	\end{aligned}
\end{equation}
which are here expressed in both the time domain and frequency domain. The frequency domain expressions are computationally efficient to calculate; however, the time domain expressions are also shown here for clarity. By maximizing $\ln{\mathcal{L}(\boldsymbol{\mu}(s_0,t_{\rm{s}},\boldsymbol{\theta});\boldsymbol{x},\boldsymbol{\Sigma})}$ with respect to $s_0$,
\begin{equation}
	\begin{aligned} \label{eq_maximize_over_amplitude}
        & \frac{\partial}{\partial s_0} \bigg(- \frac{1}{2}  \big(\boldsymbol{x}-s_0 \boldsymbol{\mu_0}(t_{\rm{s}},\boldsymbol{\theta}))\big)^\mathrm{T} \boldsymbol{\Sigma}^{-1}  \big(\boldsymbol{x}-s_0 \boldsymbol{\mu_0}(t_{\rm{s}},\boldsymbol{\theta}))\big) \bigg) \\ & = 0,
	\end{aligned}
\end{equation}
we get 
\begin{equation}
	\begin{aligned}
		\hat{s}_0(t_{\rm{s}}, \boldsymbol{\theta}) & = \frac{y_\text{mf}(t_{\rm{s}}, \boldsymbol{\theta})}{y_{\boldsymbol{\mu}}(\boldsymbol{\theta})},
	\end{aligned}
\end{equation}
which is the well-known matched filter estimate of $s_0$. By normalizing the matched filter output with the template normalization factor, $\sigma(\boldsymbol{\theta}) = y_{\boldsymbol{\mu}}(\boldsymbol{\theta})^{-1/2}$, the SNR of the template can be calculated as follows:
\begin{equation} \label{eq_mf_snr}
    \text{SNR}_\text{mf}(t_{\rm{s}},\boldsymbol{\theta}) = \frac{|y_\text{mf}(t_{\rm{s}}, \boldsymbol{\theta})|}{\sqrt{y_{\boldsymbol{\mu}}(\boldsymbol{\theta})}}.
\end{equation}
A \textit{matched filter search} is performed by calculating $\text{SNR}_\text{mf}(t_{\rm{s}},\boldsymbol{\theta})$ for one or more templates for all possible $t_{\rm{s}}$. By selecting the $t_{\rm{s}}$ and $\boldsymbol{\theta}$ that maximizes $\text{SNR}_\text{mf}(t_{\rm{s}},\boldsymbol{\theta})$, the best matching template and time is found, and $\text{SNR}_\text{mf}(t_{\rm{s}},\boldsymbol{\theta})$ serves as a significance for the signal, i.e., if $\text{max}\big(\text{SNR}_\text{mf}(t_{\rm{s}},\boldsymbol{\theta})) \big) \gg 1$, the data likely contains a signal consistent with the template. The matched filter quantities can be related to the likelihood by~\cite{Zubeldia:2021mzx}:
\begin{equation} \label{eq_likelihood_matched_filter}
\begin{aligned}
    \ln \mathcal{L}(t_{\rm{s}}, & \boldsymbol{\theta};\boldsymbol{x},S_n) \big\vert_{\hat{s}_0(t_{\rm{s}})} = -\frac{1}{2} y_{\boldsymbol{x}} + \frac{1}{2} \frac{|y_\text{mf}(t_{\rm{s}}, \boldsymbol{\theta})|^2}{y_{\boldsymbol{\mu}}(\boldsymbol{\theta})}.
\end{aligned}
\end{equation}
Since the matched filter likelihood has been analytically maximized over the signal amplitude in Eq.~\eqref{eq_maximize_over_amplitude} and is easy to scan for all possible times, it is an extremely computationally efficient way to calculate the maximum likelihood (over time and amplitude) for an already simulated signal template.

As we demonstrate in Sec.~\ref{sec_forward_folding_likelihood}, the matched filter can be used in a likelihood-based reconstruction where the parameters $\boldsymbol{\theta}$ are treated as free parameters. This eliminates two of the parameters from the reconstruction, namely the time and amplitude, and can improve the stability and speed of a reconstruction algorithm. In the case of a radio neutrino signal, $t_0$ represents the interaction time and $s_0$ corresponds to the energy of the particle shower.

The matched filter can be generalized to more antennas by adding a sum over the antennas, e.g., in Eq.~\eqref{eq_mf_snr}, and having the trace, signal, and noise power spectral density per antenna. The relative time delay of the signal should be included in the signal templates, and the time $t_0$ is then the global time offset of the signal in all antennas. In the frequency domain, the most efficient implementation can be achieved by flattening the traces, signals, and noise power spectral densities to a signal vector, and letting $k$ run from $1$ to $n_\text{antennas} \cdot n_t / 2$ to avoid the sum over antennas. This optimized matched filter method is implemented in NuRadioReco.

\section{Biases of reconstructed in-ice radio neutrino parameters} \label{sec_neutrino_reco_biases}

\begin{table*}[t]
    \centering
\begin{tabular}{|c|c||c|c|c|c|c|c|c|} 
\hline
Event & Method & $\psi_{E_\nu}$ & $\psi_{\theta_\nu}$ & $\psi_{\phi_\nu}$ & $\psi_{r_\text{vertex}}$ & $\psi_{\theta_\text{vertex}}$ & $\psi_{\phi_\text{vertex}}$ & $\psi_{t_0}$ \\ \hline \hline
1
& \begin{tabular}[c]{@{}c@{}} $\mathcal{L}$ \\  $\chi^2$\end{tabular}
& \begin{tabular}[c]{@{}r@{}} (1.5$\pm$0.3)\% \\ (2.1$\pm$0.3)\% \end{tabular}
& \begin{tabular}[c]{@{}r@{}} (0.13$\pm$0.03)$^{\circ}$ \\ (0.17$\pm$0.05)$^{\circ}$ \end{tabular}
& \begin{tabular}[c]{@{}r@{}} (0.18$\pm$0.05)$^{\circ}$ \\ (0.22$\pm$0.06)$^{\circ}$ \end{tabular}
& \begin{tabular}[c]{@{}r@{}} (-0.1$\pm$0.2)\,m \\ (0.1$\pm$0.2)\,m \end{tabular}
& \begin{tabular}[c]{@{}r@{}} (0.034$\pm$0.034)$^{\circ}$ \\ (0.001$\pm$0.039)$^{\circ}$ \end{tabular}
& \begin{tabular}[c]{@{}r@{}} (-0.009$\pm$0.005)$^{\circ}$ \\ (-0.005$\pm$0.006)$^{\circ}$ \end{tabular}
& \begin{tabular}[c]{@{}r@{}} (-0.001$\pm$0.002)\,ns \\ (-0.001$\pm$0.002)\,ns \end{tabular}
\\ \hline
2
& \begin{tabular}[c]{@{}c@{}} $\mathcal{L}$ \\  $\chi^2$\end{tabular}
& \begin{tabular}[c]{@{}r@{}} (2.2$\pm$0.5)\% \\ (4.2$\pm$0.8)\% \end{tabular}
& \begin{tabular}[c]{@{}r@{}} (0.64$\pm$0.07)$^{\circ}$ \\ (0.68$\pm$0.08)$^{\circ}$ \end{tabular}
& \begin{tabular}[c]{@{}r@{}} (-0.33$\pm$0.39)$^{\circ}$ \\ (-0.33$\pm$0.40)$^{\circ}$ \end{tabular}
& \begin{tabular}[c]{@{}r@{}} (-3.3$\pm$3.6)\,m \\ (3.6$\pm$5.9)\,m \end{tabular}
& \begin{tabular}[c]{@{}r@{}} (0.006$\pm$0.003)$^{\circ}$ \\ (0.004$\pm$0.005)$^{\circ}$ \end{tabular}
& \begin{tabular}[c]{@{}r@{}} (0.001$\pm$0.001)$^{\circ}$ \\ (0.001$\pm$0.001)$^{\circ}$ \end{tabular}
& \begin{tabular}[c]{@{}r@{}} (-0.004$\pm$0.002)\,ns \\ (-0.005$\pm$0.002)\,ns \end{tabular}
\\ \hline
3
& \begin{tabular}[c]{@{}c@{}} $\mathcal{L}$ \\  $\chi^2$\end{tabular}
& \begin{tabular}[c]{@{}r@{}} (3.4$\pm$0.5)\% \\ (3.3$\pm$0.6)\% \end{tabular}
& \begin{tabular}[c]{@{}r@{}} (0.21$\pm$0.03)$^{\circ}$ \\ (0.31$\pm$0.03)$^{\circ}$ \end{tabular}
& \begin{tabular}[c]{@{}r@{}} (-0.90$\pm$0.17)$^{\circ}$ \\ (-1.21$\pm$0.21)$^{\circ}$ \end{tabular}
& \begin{tabular}[c]{@{}r@{}} (1.5$\pm$0.9)\,m \\ (-0.6$\pm$1.1)\,m \end{tabular}
& \begin{tabular}[c]{@{}r@{}} (-0.003$\pm$0.044)$^{\circ}$ \\ (0.124$\pm$0.051)$^{\circ}$ \end{tabular}
& \begin{tabular}[c]{@{}r@{}} (0.004$\pm$0.006)$^{\circ}$ \\ (-0.010$\pm$0.007)$^{\circ}$ \end{tabular}
& \begin{tabular}[c]{@{}r@{}} (-0.004$\pm$0.005)\,ns \\ (-0.007$\pm$0.005)\,ns \end{tabular}
\\ \hline
\end{tabular}
\caption{Biases, $\psi_\theta = E[\theta_{\text{reco}} - \theta_{\text{true}}]$, of reconstructed signal parameters for the three neutrino events in Table~\ref{tab_events} with 500 different realizations of noise each using the $-2\ln{\mathcal{L}}$ or the $\chi^2$ as the objective function in the reconstruction. The bias on the shower energy is given as a percentage of the true shower energy. The statistical uncertainties on the biases are the errors of the means, i.e., the spreads (Table~\ref{table_llh_vs_chi2}) divided by the square root of the number of events.}
\label{table_llh_vs_chi2_biases}
\end{table*}

To further assess the importance of considering noise correlations in reconstruction of radio neutrino signals (as discussed in Sec.~\ref{sec_chi2_comparison}), we quantify the biases of the reconstruction methods using the $-2\ln{\mathcal{L}}$ or $\chi^2$ as objective functions. We define the bias as $\psi_\theta = E[\theta_{\text{reco}} - \theta_{\text{true}}]$, which we calculate using 500 reconstructions of the same three neutrino events with different realizations of noise. The resulting biases for each neutrino parameter using the two methods is shown in Table~\ref{table_llh_vs_chi2_biases}, including their statistical uncertainties (error of mean). In general, both methods have small biases, often on the order of $0.2 \sigma$ compared to the uncertainties in Table~\ref{table_llh_vs_chi2}. For the parameters of most importance, the neutrino energy and arrival direction, the likelihood method has smaller biases than the $\chi^2$ method. For the parameters related to the vertex position, the results are less conclusive due to statistical uncertainties.

\bibliographystyle{JHEP}
\bibliography{refs}

\end{document}